\documentclass[11pt,a4paper]{article}
\pdfoutput=1

\usepackage{amsmath,amsfonts,amsbsy,amssymb,array,accents,dsfont}
\usepackage{enumerate,array,latexsym,graphicx,mathrsfs,verbatim,psfrag}
\usepackage{bm} 
\usepackage{booktabs} 
\usepackage[usenames]{color}
\usepackage{setspace}
\usepackage[english]{babel}
\usepackage{fancybox}
\usepackage{datetime}
\usepackage[nosort]{cite}
\usepackage{chngpage} 
\usepackage{gensymb} 
\usepackage{wasysym}
\usepackage{psfrag}
\usepackage{epstopdf}  

\usepackage{enumitem}
\setlist{itemsep=0pt}

\usepackage{mciteplus}

\usepackage[colorlinks=true,      linkcolor=darkblue,      urlcolor=darkblue,      
            filecolor=darkblue,      citecolor=darkblue,       pdfstartview=FitH,     
						pdfpagemode=UseNone,      bookmarksopen=true]{hyperref}  
\usepackage[all]{hypcap}     

\def\eq#1{(\ref{#1})}

\newcommand{\captionfonts}{\small}
\makeatletter  
\long\def\@makecaption#1#2{%
  \vskip\abovecaptionskip
  \sbox\@tempboxa{{\captionfonts #1: #2}}%
 \ifdim \wd\@tempboxa >\hsize
    {\captionfonts #1: #2\par}
  \else
    \hbox to\hsize{\hfil\box\@tempboxa\hfil}%
  \fi
  \vskip\belowcaptionskip}
\makeatother   

\DeclareMathSymbol{\medhatsym}{\mathord}{largesymbols}{"62} 
\newcommand\lowermedhatsym{
  \text{\smash{\raisebox{-1.28ex}{%
    $\medhatsym$}}}}
\newcommand\medhat[1]{
  \mathchoice
    {\accentset{\displaystyle\lowermedhatsym}{#1}}
    {\accentset{\textstyle\lowermedhatsym}{#1}}
    {\accentset{\scriptstyle\lowermedhatsym}{#1}}
    {\accentset{\scriptscriptstyle\lowermedhatsym}{#1}}
}

\DeclareMathSymbol{\medtildesym}{\mathord}{largesymbols}{"65}
\newcommand\lowermedtildesym{
  \text{\smash{\raisebox{-1.2ex}{%
    $\medtildesym$}}}}
\newcommand\medtilde[1]{
  \mathchoice
    {\accentset{\displaystyle\lowermedtildesym}{#1}}
    {\accentset{\textstyle\lowermedtildesym}{#1}}
    {\accentset{\scriptstyle\lowermedtildesym}{#1}}
    {\accentset{\scriptscriptstyle\lowermedtildesym}{#1}}
}

\newcommand{\comm}[1]{} 

\def\IR{\mathbb{R}}

\def\({\left(}
\def\){\right)}
\def\[{\left[}
\def\]{\right]}

\def\hk{hyperk\"ahler}

\def\coeff#1#2{{\textstyle \frac{#1}{#2}}}

\def\One{{\hbox{ 1\kern-.8mm l}}}

\def\barray{\begin{array}}
\def\earray{\end{array}}
\def\be{\begin{equation}}
\def\ee{\end{equation}}
\def\bea{\begin{eqnarray}}
\def\eea{\end{eqnarray}}
\def\bal{\begin{align}}
\def\eal{\end{align}}

\numberwithin{equation}{section} 

\makeatletter
\g@addto@macro\bfseries{\boldmath}
\makeatother

\definecolor{cardinal}{rgb}{0.6,0,0}
\definecolor{darkgreen}{rgb}{0,0.4,0}
\definecolor{purple}{rgb}{0.5, 0, 0.5}
\definecolor{golden}{rgb}{0.92, 0.7, 0}
\definecolor{midnight}{rgb}{0, 0, 0.5}
\definecolor{darkblue}{rgb}{0, 0, 0.7}

\def\coeff#1#2{\relax{\textstyle {#1 \over #2}}\displaystyle}

\def\IR{\mathbb{R}}

\def\cA{{\cal A}}

\def\cD{{\cal D}}

\def\cL{{\cal L}}
\def\cK{{\cal K}}
\def\cM{{\cal M}}
\def\cN{{\cal N}}
\def\cO{{\cal O}}
\def\cP{{\cal P}}

\def\cO{{\cal O}}

\newcommand{\adstwo}{\ensuremath{\mathrm{AdS}_2}}
\newcommand{\adsthree}{\ensuremath{\mathrm{AdS}_3}}
\newcommand{\adsfour}{\ensuremath{\mathrm{AdS}_4}}

\newcommand{\kk}{\ensuremath{\boldsymbol{\omega}}}
\newcommand{\K}{\cK}

\newcommand{\sqrtpadstwo}{\ensuremath{\sqrt{\medtilde{\cP}\phantom{{\medhat{\Sigma}}^{(9)}}\hspace{-6.8mm}}}}
\newcommand{\sigmaadstwodenom}{\ensuremath{\medtilde{\Sigma}\phantom{{\medhat{\Sigma}}^{(9)}}\hspace{-6.8mm}}}
\newcommand{\tildespace}{\ensuremath{\phantom{{\medhat{\Sigma}}^{(9)}}\hspace{-6.8mm}}}
\newcommand{\rt}{\ensuremath{\tilde{r}}}
\newcommand{\Ft}{\ensuremath{\medtilde{F}}}

\newcommand{\T}[3]{\ensuremath{ #1{}^{#2}_{\phantom{#2} \! #3}}}		

\def\vh{\hat{v}}

\newcommand{\brho}{\ensuremath{\boldsymbol{\rho}}}

\begin{document}

\phantom{AAA}
\vspace{-10mm}

%
%

\vspace{33mm}

\begin{center}

{\huge {\bf AdS$_2$ Holography: Mind the Cap}}

\bigskip

\vspace{15mm}

{\large
\textsc{Iosif Bena$^1$,~ Pierre Heidmann$^{1}$,~ David Turton$^{2}$}}

\vspace{12mm}

$^1$Institut de Physique Th\'eorique,\\
Universit\'e Paris Saclay,\\
CEA, CNRS, F-91191 Gif sur Yvette, France \\
\bigskip
$^2$Mathematical Sciences and STAG Research Centre, \\ University of Southampton,\\ Highfield,
Southampton SO17 1BJ, United Kingdom\\

\vspace{9mm} 
{\footnotesize\upshape\ttfamily iosif.bena @ ipht.fr, ~pierre.heidmann @ ipht.fr, ~d.j.turton @ soton.ac.uk} \\

\vspace{25mm}
 
\textsc{Abstract}

\end{center}

\begin{adjustwidth}{15mm}{15mm} 
 
\vspace{1mm}
\noindent
AdS$_2$ plays an extremely important role in black-hole physics. We construct several infinite families of supergravity solutions that are asymptotically AdS$_2$ in the UV, and terminate in the IR with a cap that is singular in two dimensions but smooth in ten dimensions. These solutions break conformal invariance, and should correspond to supersymmetric ground states of a holographically dual CFT$_1$. We solve the free massless scalar wave equation on a family of these solutions, finding towers of finite-energy normalizable bound-state excitations. We discuss the intriguing possibility that these excitations correspond to time-dependent excitations of the dual CFT$_1$, which would imply that this CFT$_1$ is dynamical rather than topological, and hence cannot have a conformally invariant ground state.

\vspace{2.8cm}

\centerline{\it Dedicated to the memory of Joe Polchinski}

\end{adjustwidth}

\thispagestyle{empty}
\newpage


\baselineskip=14.7pt
\parskip=3pt

\setcounter{tocdepth}{2}
\tableofcontents

\baselineskip=15pt
\parskip=3pt



\section{Introduction}

Understanding quantum gravity in spacetimes that asymptote to two-dimensional Anti-de Sitter space (\adstwo) times a compact manifold $\K$ is one of the most challenging and interesting problems in theoretical physics at present, for three reasons. First, string theory has had great success in counting the microstates of extremal black holes whose near-horizon geometries contain a factor that is \adsthree\  \cite{Strominger:1996sh,Sen:1995in,Maldacena:1997de} or \adsfour\ \cite{Benini:2016rke,Azzurli:2017kxo}, however many extremal black holes have an \adstwo\ near-horizon limit that is not contained in a higher-dimensional AdS space, and the counting of the microstates of these black holes is poorly understood. Furthermore, many black holes have an \adsthree\ near-horizon limit and a further \adstwo\ very-near-horizon geometry deeper in the infrared. For these black holes, understanding the RG flow between \adsthree\ and \adstwo\ remains an important and challenging open problem \cite{Strominger:1998yg,Gupta:2008ki,Balasubramanian:2009bg,*Balasubramanian:2010ys,Castro:2014ima,Cvetic:2016eiv}.

Second, holography in \adstwo\ is somewhat subtle: it is well known that the backreaction of finite-energy excitations in global \adstwo\ necessarily diverges at one of the two asymptotic boundaries~\cite{Maldacena:1998uz,Almheiri:2014cka}. Indeed, much of the recent interest in the Sachdev-Ye-Kitaev (SYK) model and its dual (see for example \cite{Kitaev:2015talks,Maldacena:2016hyu,*Cotler:2016fpe,*Balasubramanian:2016ids,*Kitaev:2017awl}) is driven by the desire to understand quantum gravity in \adstwo. Since global \adstwo\ has two disconnected boundaries, it appears that its holographic dual should be two copies of a CFT$_1$.\footnote{See \cite{Bena:2007ju,Lunin:2015hma} for work on the construction of bubbling solutions that are asymptotic to global \adstwo.} By contrast, black hole entropy in String Theory is usually accounted for by enumerating bound states of a (single) system of branes, so one expects there to be an AdS$_2$/CFT$_1$ entropy calculation that involves counting ground states of a single CFT$_1$~(see for example \cite{Gupta:2008ki,Sen:2011cn}). 
It does not appear to be understood in general whether the ground states of the CFT$_1$ preserve or break conformal invariance, whether the CFT$_1$ is topological, and whether or not one can construct a tower of non-supersymmetric states above a given ground state. 
When the \adstwo\ is in the infrared of an \adsthree\ space, and hence the CFT$_1$ has a CFT$_2$ parent theory, the states of the CFT$_1$ have also been argued to have a description as the states of the chiral half of this CFT$_2$, at least in certain contexts~\cite{Boonstra:1998yu,Strominger:1998yg,Gupta:2008ki,Balasubramanian:2009bg,*Balasubramanian:2010ys}. However, in most duality frames supersymmetric black holes do not have such a parent \adsthree/CFT$_2$.

Third, there are several arguments that at the horizon of all black holes there should be some structure that allows information to escape in order to render black hole evaporation a unitary quantum process \cite{Mathur:2005zp, Mathur:2009hf, Almheiri:2012rt}. 
Much of this structure has been constructed so far in the context of the microstate geometry programme \cite{Bena:2006kb,Bena:2007qc,Bena:2010gg,Bena:2016ypk,Bena:2017xbt}, 
where it is understood how this structure avoids collapsing into a black hole \cite{Gibbons:2013tqa,*deLange:2015gca,*Haas:2014spa}. It has been argued \cite{Dabholkar:2010rm,Chowdhury:2015gbk, Chowdhury:2014yca} that requiring that a solution have black hole asymptotics does not guarantee that this solution is a true black hole microstate. Four-dimensional supersymmetric black holes have an \adstwo\ near-horizon region and zero angular momentum; in five dimensions there are two angular momenta,  $J_L$, $J_R$, and the supersymmetric BMPV~\cite{Breckenridge:1996is} black hole has $J_R=0$. For such black holes, according to~\cite{Dabholkar:2010rm,Chowdhury:2015gbk}, the only solutions that can be interpreted as pure black hole microstates (involving no additional degrees of freedom exterior to the black hole) are those that, modulo some dressing with a small number of fermion zero modes, have zero angular momentum in 4D ($J_R=0$ in 5D) and fit in an \adstwo\ region. 

The purpose of this paper is to address all these three points at the same time, by constructing families of smooth solutions that have an $\adstwo\times \K$ asymptotic region in the UV (allowing also for $\K$ to be non-trivially fibered over \adstwo), that end in the IR with a smooth cap, and that have $J_R=0$. 

According to the AdS-CFT correspondence, these supersymmetric solutions should be dual to pure states of (a single copy of) a CFT$_1$. Furthermore, our new six-dimensional solutions are obtained by taking a scaling limit of a class of asymptotically $\adsthree\times $S$^3$ solutions with identified dual CFT$_2$ states, providing an implicit map between the states of this CFT$_1$ and the states of the parent D1-D5 CFT.

Each of the supersymmetric solutions we construct caps off smoothly in its deep interior. This capping off involves the smooth shrinking of a cycle in $\cK$ which locally becomes the angular direction of polar coordinates of an $\IR^2$ factor of the local geometry. As a result, if one reduces these solutions to two-dimensional gravity coupled to matter, they appear to be geometrically singular. However, this singularity is resolved into smooth geometry supported by fluxes in five- or six-dimensional supergravity (and ultimately into ten- or eleven-dimensional supergravity). 

The naive two-dimensional geometrical singularity of our solutions enables our solutions to have non-trivial features in their IR while preserving the \adstwo\ UV, in contrast to the rigidity of global AdS$_2$~\cite{Maldacena:1998uz,Almheiri:2014cka}. 
It is tempting to think of our solutions as corresponding to string-theoretical resolutions of the IR singularity arising from the backreaction of finite-energy excitations in global AdS$_2$, although it should be said that it is far from obvious that our solutions arise in this fashion.
Nevertheless, pursuing the qualitative analogy, this would be analogous to how the Polchinski-Strassler brane polarization, visible only in ten dimensions \cite{Polchinski:2000uf}, resolves the singularity of the five-dimensional gauged supergravity flow of GPPZ \cite{Girardello:1998pd} (note also the very recent work on constructing the 10D supergravity uplift of the GPPZ solution~\cite{Petrini:2018pjk,*Bobev:2018eer,*Bena:2018vtu}).

By contrast, most of the recent attempts to understand quantum gravity in \adstwo\ involve modifying the UV (by adding a running dilaton and working in a ``Nearly-\adstwo'' geometry) and preserving the IR~\cite{Almheiri:2014cka,Maldacena:2016hyu,*Cotler:2016fpe,*Balasubramanian:2016ids,*Kitaev:2017awl}. Indeed, if one works in two-dimensional theories with relatively simple field content, modifying the UV is the only option; our solutions require much richer field content from a two-dimensional perspective, as of course is natural in String Theory.

These two options: either keeping the UV fixed and resolving the IR singularity by brane polarization and bubbling (as in Polchinski-Strassler) or keeping the IR fixed and modifying the UV asymptotics (as in Almheiri-Polchinski), appear to be the only two possibilities to obtain non-trivial physics in \adstwo. Our interest in the present work is in the CFT$_1$ description of asymptotically AdS$_2$ string theory solutions, rather than irrelevant deformations of such a CFT$_1$. We therefore choose the Polchinski-Strassler option over the Almheiri-Polchinski one.

The presence of a smooth IR cap allows our supersymmetric solutions to support an infinite tower of non-supersymmetric linearized excitations. These excitations are localized very near the IR cap, and are normalizable. 
An important question that remains open is whether or not the backreaction of these excitations preserves the \adstwo\ UV asymptotics.
One possibility, consistent with the naive extrapolation of the results of~\cite{Maldacena:1998uz,Almheiri:2014cka} to our capped solutions, is that the backreaction of these excitations necessarily modifies the \adstwo\ UV asymptotics, meaning that these excitations are not dual to any states of the original CFT$_1$. 
However, since there exist non-supersymmetric black hole solutions with a finite bulk stress-energy tensor that preserve the \adstwo\ asymptotics \cite{Castro:2008ms}, it is possible that the backreacted non-supersymmetric solutions will also preserve the \adstwo\ asymptotics, and indeed this is our expectation. 
We rather expect that the data that determines whether the UV is modified is independent of the existence of our excitations, and will discuss this in more detail in Section~\ref{subsec:backreaction}. 

If the backreaction of our time-dependent perturbations preserves the AdS$_2$ UV asymptotics, then these perturbations should be dual to time-dependent excitations of the CFT$_1$. This in turn indicates that this CFT$_1$ has nontrivial dynamics, and hence is not a topological theory. By contrast, if a given CFT$_1$ has a conformally invariant ground state (which would presumably be holographically dual to empty Poincar\'e \adstwo), it is necessarily topological.\footnote{We thank Miguel Paulos for communicating to us this statement and a supporting argument.} Hence, if the backreaction preserves the \adstwo\ UV, this implies that the dual CFT$_1$ does not have a conformally invariant ground state. CFT$_1$ models that have no conformally invariant ground state have been discussed in~\cite{Chamon:2011xk,*Jackiw:2012ur}. A related interesting question is whether or not the backreacted time-dependent solutions have zero holographic stress tensor (as the constant-dilaton solutions analyzed in~\cite{Cvetic:2016eiv}), or rather whether their time-dependence implies they correspond to finite-energy states of the CFT$_1$; we will discuss this question in detail in Section~\ref{sec:disc-holo}.

The absence of a conformally invariant ground state would in turn indicate that empty Poincar\'e \adstwo\ cannot be dual to any pure state of the microscopic CFT$_{1}$ under consideration. From this perspective Poincar\'e \adstwo\  would have a similar status to Poincar\'e \adsthree, which is also not dual to any pure state of the D1-D5 CFT, but is a singular geometry that should be rather thought of as an approximation to a mixed state. The bulk duals of all the pure states of a single  CFT$_{1}$ would therefore be asymptotically \adstwo\ states of String Theory with nontrivial (and likely stringy and/or quantum) physics in the infrared that breaks conformal invariance.

In the context of black hole quantum physics, the caps of our \adstwo\ solutions correspond to new physics at the scale of the would-be horizon of the black hole; such non-trivial horizon-scale physics is anticipated by the fuzzball proposal  \cite{Mathur:2005zp,Bena:2007kg,Skenderis:2008qn}. While this proposal is compatible with the existence or otherwise of a conformally invariant CFT$_1$ ground state, if one could prove that there is no such ground state, this would establish beyond reasonable doubt that the fuzzball proposal is the correct description of extremal black holes. Note that such a proof would not rely on the construction of the holographic duals to typical CFT$_1$ states, which may involve string-theoretic degrees of freedom beyond supergravity; it would be sufficient to backreact the perturbations of the somewhat atypical microstates  that are described by the smooth supergravity solutions that we construct in the present work.

The organization of this paper is as follows. In Section \ref{sec-over} we give an overview of our results and their implications. In Section \ref{sec:general-limit} we formulate our general \adstwo\ limit for smooth horizonless supergravity solutions. In Section \ref{sec3} we construct a class of asymptotically \adstwo\ multi-center solutions. In Section \ref{sec4} we transform a family of superstratum solutions with long scaling throats into asymptotically \adstwo\ solutions and discuss their description in the dual CFT$_2$. In Section \ref{sec5} we solve the free massless scalar wave equation on a class of asymptotically \adstwo\ superstrata, finding an infinite tower of bound-state excitations. Finally, in Section \ref{discussion} we discuss the backreaction of these excitations, their holographic description, and possible connections with other approaches to \adstwo\ quantum gravity.  In Appendix \ref{app:BPS-equations} we record for completeness the BPS equations in the five- and six-dimensional supergravity theories in which we work. In Appendix \ref{app:WEresolution} we detail the method we use to solve the free massless scalar wave equation on asymptotically \adstwo\ superstrata.

\section{Overview of results}
\label{sec-over}

\subsection{The construction}

We shall define a general scaling limit for smooth horizonless supergravity solutions, focusing on solutions in either five or six dimensions.

The five-dimensional solutions that we start with are asymptotically $\IR^{4,1}$, with a throat that is approximately that of the near-horizon BMPV solution~\cite{Breckenridge:1996is}. (We remind the reader that this solution can be written as S$^3$ fibered over \adstwo, and that when $J_L=0$ the fibration becomes trivial and the solution is \adstwo$\times$S$^3$.) All the solutions we consider cap off smoothly in their core. The scaling limit transforms these solutions into asymptotically near-horizon-BMPV solutions with smooth caps.

The six-dimensional capped solutions that we start with 
are asymptotically either $\IR^{4,1}\times$S$^1$ or AdS$_3\times$S$^3$, with a throat that is S$^3$ fibered over the extremal BTZ black hole~\cite{Banados:1992wn}. In the ``very-near-horizon'' limit this can written as S$^1$ fibered over the near-horizon BMPV solution (see for example~\cite{Strominger:1998yg}). Our scaling limit similarly transforms these solutions into solutions whose asymptotics are S$^1$ fibered over near-horizon BMPV, with smooth caps. We shall present an explicit family of examples where the throat is S$^1$ fibered over AdS$_2\times$S$^3$. For ease of language, when discussing both five- and six-dimensional solutions we shall often speak in terms of \adstwo\ limits, \adstwo~throats, and asymptotically \adstwo\ solutions.

\begin{figure}[hbt!]
\centering
\vspace{3mm}
\includegraphics[width=117mm]{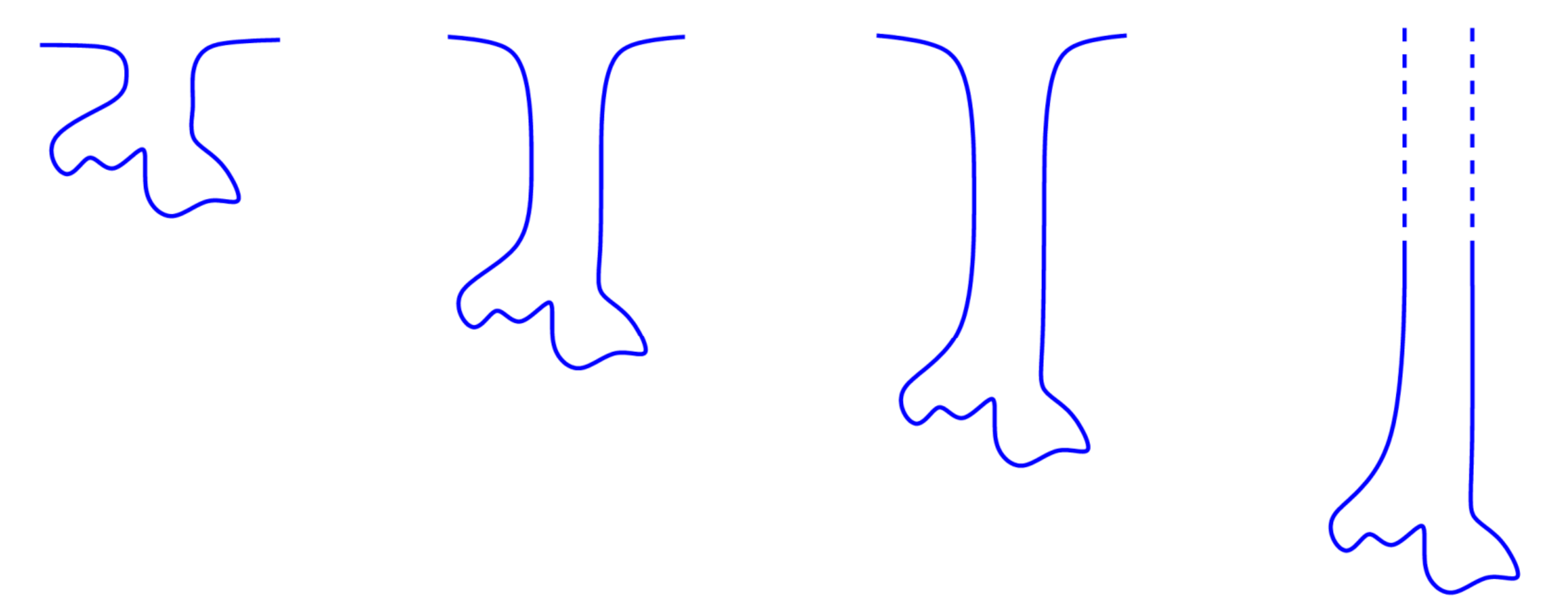}
\vspace{2mm}
\caption{A schematic pictorial representation of a scaling geometry and the asymptotically \adstwo\ limit. The proper size of the bubbles remains the same as the throat of the solutions becomes longer~\cite{Bena:2006kb,Bena:2007qc}; the throat becomes infinite in the \adstwo\ limit.}
\label{ScalingProcess}
\vspace{0mm}
\end{figure}

These five- and six-dimensional limits correspond to the two classes of deep scaling microstate solutions that are known so far.
The first class comprises supersymmetric solutions of five-dimensional U(1)$^n$ ungauged supergravity, that have a Gibbons-Hawking four-dimensional base space and preserve the same tri-holomorphic isometry as this space. 
These solutions are determined by $2n+2$ harmonic functions in the $\IR^3$ base of the Gibbons-Hawking space \cite{Gutowski:2004yv,*Gauntlett:2004qy,*Bena:2004de,Bena:2005ni}; the poles of the harmonic functions are arranged to avoid horizons and singularities \cite{Bena:2005va,Berglund:2005vb}. These solutions have nontrivial bubbles (two-cycles) threaded by fluxes, and have the same mass, charges and angular momenta as black holes with a macroscopically large horizon area. Furthermore, one can construct families of ``scaling'' solutions \cite{Bena:2006kb,Bena:2007qc,Denef:2002ru}, where by adjusting a parameter, the $\IR^3$ distance between the poles of the harmonic functions can be made arbitrarily small, whereupon the solution develops an arbitrarily long \adstwo\ throat. In the full five- or six-dimensional solution, the proper size of the bubbles nevertheless remains fixed during the scaling process (see Fig.\;\ref{ScalingProcess})~\cite{Bena:2006kb,Bena:2007qc}.

The first deep scaling solutions with a Gibbons-Hawking base were constructed in a somewhat artisanal fashion \cite{Bena:2006kb}, however a systematic methodology for building such solutions has recently been developed \cite{Heidmann:2017cxt}. This has resulted in the construction of the largest known classes of such solutions \cite{Heidmann:2017cxt,Bena:2017fvm,Avila:2017pwi}. We will show that all these solutions admit a scaling limit that amounts to setting to zero all the constants in the harmonic functions. For scaling solutions this can be done without introducing closed timelike curves, and results in asymptotically \adstwo\ geometries whose infrared ends in a smooth cap.

The second class of deep scaling solutions is more recent~\cite{Bena:2016ypk,Bena:2017xbt}, and these solutions preserve fewer isometries than the solutions described above. They are smooth in supergravity only in certain duality frames, such as the frame where their charges correspond to D1 branes, D5 branes and momentum P.  These solutions are constructed using ``superstratum'' technology~\cite{Bena:2016ypk,Bena:2017xbt,Lunin:2012gp,Giusto:2013bda,Bena:2015bea,Bena:2016agb,Bena:2017geu}, and are parameterized by arbitrary functions of at least two variables. 

The superstratum solutions have two important advantages that justify the use of the more complicated technology.
The first advantage is that in an appropriate regime of parameters these solutions have an intermediate approximately $\adsthree\times$S$^3$ region; one can decouple this region and its interior from the flat asymptotics in the usual way. In the \adsthree\ decoupling limit, these solutions correspond to families of states in the dual CFT$_2$; these states have been explicitly identified at the orbifold locus in moduli space~\cite{Bena:2016ypk,Bena:2017xbt}. In contrast, while the Gibbons-Hawking multicenter bubbling solutions can also have an \adsthree\  region, no holographic description of these solutions is known. The second advantage of the superstratum solutions over the five-dimensional bubbling solutions is that one has explicit parametric control over both the angular momenta.

The resulting asymptotically \adstwo\ superstrata are parameterized by the same number of functions as the original superstrata. Hence, our construction provides a map between the CFT$_{2}$ states dual to asymptotically \adsthree\ superstrata and the states of the CFT$_{1}$ dual to the corresponding asymptotically \adstwo\ solutions. Furthermore, if the counting of superstrata eventually matches (an order-one fraction of) the entropy of the corresponding black hole, as suggested in \cite{Bena:2014qxa}, one would expect that the counting of the asymptotically \adstwo\ solutions constructed from them will also match this entropy. A definitive answer to this question is beyond the scope of this paper.

\subsection{The \adstwo--\adsthree\ connection and black hole microstate geometries}
\label{sec:ads2-ads3-overview}

Our map also clarifies the relation between deep scaling microstate geometries, the angular momentum $J_R$, and the discussions of~\cite{Dabholkar:2010rm,Chowdhury:2015gbk}. In particular, the key argument of~\cite{Dabholkar:2010rm,Chowdhury:2015gbk} was that all the information characterizing the microstates of supersymmetric black holes should fit inside an \adstwo\ throat, and everything that does not do so represents degrees of freedom external to the black hole horizon. 
Our construction shows that all the information about the topology, fluxes and wiggles of the scaling black hole microstates constructed thus far passes this criteria. The only information that does not survive the \adstwo\ limit is the non-zero $J_R$ of the asymptotically \adsthree\ solutions, which is proportional to the inverse of the length of the \adstwo\ throat and thus vanishes in the \adstwo\ scaling limit, consistent with the discussions 
in~\cite{Dabholkar:2010rm,Chowdhury:2015gbk}.
The fact that these solutions fit inside an \adstwo\ region, and that in this limit the angular momentum $J_R$ vanishes, indicates that the non-zero $J_R$ of the corresponding asymptotically \adsthree\ solutions does not come from the structure that replaces the horizon, but rather from the gluing of the long \adstwo\ throat to the ambient spacetime.

It is interesting to examine the \adstwo\ scaling procedure from an \adsthree\ perspective. As the \adstwo\ throat of the solutions becomes longer, the $J_R$ angular momentum decreases. For multi-center bubbling solutions, even if the lengthening of the throat looks continuous from a classical perspective, it is not if one includes quantum effects \cite{deBoer:2008zn,Bena:2007qc}. Similarly, for superstratum solutions $J_R$ appears to be a continuous parameter, but we know that in order for the solution  to make sense  $J_R$ should be quantized in half-integer units. Hence, both generic scaling bubbling solutions and superstrata with increasingly long throats are expected to be dual to a family of CFT$_{2}$ microstates with progressively smaller values of $J_R$. From the \adsthree\ perspective the superstratum with $J_R=1/2$ is at the extreme end of this family, and there is no superstratum with $J_R=0$. This makes perfect sense, since taking the  $J_R \rightarrow 0$ limit within this family makes the throat infinite and, from an \adsthree\ perspective, produces the classical black hole solution~\cite{Bena:2016ypk}.
 
However from the perspective of an observer at the bottom of the \adstwo\ throat, the physics is very different. As one lowers the value of $J_R$, the \adstwo\ throat becomes longer and longer, so this observer sees an increasingly large region of \adstwo\ that connects far away to an \adsthree\ or to a flat region. (From an \adsthree\ perspective this observer has lower and lower energies as one increases the length of the throat.) As $J_R$ is taken to zero, the locus of the connection of the \adstwo\ region with the \adsthree\ or the flat region is taken to infinite distance, and one obtains an asymptotically \adstwo\ throat that caps off smoothly in the infrared. However, during this process the local physics seen by the observer at the bottom of the throat remains essentially the same, since the region where the \adstwo\ is glued with the asymptotic space is far away.

\subsection{Quantum Gravity and String Theory in \adstwo}

Our results have implications for other efforts to understand quantum gravity in \adstwo. 
In contrast to our solutions with non-trivial IR physics, most of the 
work that has recently taken place on \adstwo\ and Nearly-\adstwo\ quantum gravity, 
including the interest in the SYK model and related models 
(see for example~\cite{Kitaev:2015talks,Maldacena:2016hyu,*Cotler:2016fpe,*Balasubramanian:2016ids,*Kitaev:2017awl}), 
involves deforming the UV. For example, Jackiw-Teitelboim 
gravity \cite{Jackiw:1984je,*Teitelboim:1983ux} has been argued to capture 
the universal sector of excitations in Nearly-\adstwo\ 
backgrounds~\cite{Almheiri:2014cka,Maldacena:2016upp,Jensen:2016pah,Engelsoy:2016xyb}. 
These excitations are localized in the region of the gluing between \adstwo\ and the 
UV geometry, and are described by the Schwarzian action; they appear to be similar 
to the singleton modes of the AdS$_5$/CFT$_4$ 
correspondence~\cite{Witten:1998qj,*deBoer:1998ip}, and related modes in 
\adsthree~\cite{Brown:1986nw,Mathur:2011gz,*Mathur:2012tj,Lunin:2012gp,Giusto:2013bda}. 
As such, we do not expect this action to describe the full dynamics of the CFT$_1$, 
much as the singleton modes do not capture the SU(N) part of the dynamics of 
U(N) ${\mathcal N} =4$ Super-Yang-Mills theory. Our construction supports this 
intuition by demonstrating that the details that distinguish various supersymmetric
 ground states of the CFT$_1$ are encoded in the bulk by physics in the deep IR region, 
rather than in the UV.

The fact that our capped asymptotically \adstwo\ solutions admit finite-energy bound-state
excitations reinforces this picture.
In Section \ref{discussion} we will argue that the backreaction of 
these excitations most likely preserves the \adstwo\ UV asymptotics, 
which would imply that the dual CFT$_1$ cannot have a conformally invariant 
ground state. Note that this would be a very strong statement. In particular, 
it would imply that if a CFT$_1$ proposed to describe quantum gravity in 
\adstwo\ does have a conformally invariant ground state, this theory is not 
in the same category as the 1D CFTs holographically dual to the \adstwo\ 
solutions that one constructs in String Theory. 
This would also give an explanation for the fact that backreaction in global \adstwo\ 
gives rise to singularities; un-capped \adstwo\ can only be dual to a conformally invariant 
ground state of a topological CFT$_1$.
If the actual microscopic 1D CFTs that arise in String Theory are not topological, 
statements about backreaction in un-capped \adstwo\ should not have any meaning in these theories.

The existence of an infinite tower of excited states in the CFT$_1$ dual to \adstwo\ 
appears to be in tension with arguments that this CFT$_1$ is obtained by taking 
a limit of the D1-D5 CFT that freezes all its right-moving excitations, and hence should have no 
finite-energy excitations (see for example~\cite{Strominger:1998yg,Balasubramanian:2009bg}). 
The \adstwo-\adsthree\ connection discussed in the previous subsection and the 
construction in Section \ref{sec5} clarifies this apparent tension: 
if one considers the finite-energy excitations from the perspective 
of \adsthree, these excitations are localized at the bottom of the \adstwo\ throat, 
and hence their energy becomes smaller and smaller as one makes the \adstwo\ throat 
longer and longer. However, from the the perspective of an \adstwo\ 
observer, the energy of these excitations remains finite as one takes the scaling limit, 
and hence these excitations survive. 

We now proceed to our construction, before returning to discuss the implications of our results in Section \ref{discussion}.

\section{AdS$_2$ limit of solutions of 5D and 6D supergravity}
\label{sec:general-limit}

\subsection{BPS solutions and throat geometries in 5D}

We now formulate a procedure to obtain asymptotically \adstwo\ (in general asymptotically near-horizon BMPV) solutions starting from deep scaling BPS microstate geometries of five-dimensional supergravity with $\IR^{4,1}$ asymptotics. This class of solutions is less general than the BPS solutions of six-dimensional supergravity that we shall discuss later in this section, since to reduce to five dimensions one needs to impose an extra isometry. However, their advantage is that these asymptotically $\IR^{4,1}$ solutions are smooth in any duality frame in which the horizon of the corresponding black hole is large. Hence, the resulting five-dimensional asymptotically \adstwo\ solutions describe geometric microstates of extremal black holes in all duality frames. 

Supersymmetric solutions to five-dimensional supergravity can locally be written as fibrations on a four-dimensional \hk\ base space~\cite{Gauntlett:2002nw}.
We will be interested in minimal five-dimensional supergravity coupled to Abelian vector multiplets. For concreteness we shall focus on three vector multiplets, a $U(1)^4$ theory; it is straightforward to generalize the discussion to more general field content. The ansatz we take is as follows~\cite{Gutowski:2004yv,*Gauntlett:2004qy,*Bena:2004de,Giusto:2012gt,*Vasilakis:2012zg,*Bena:2013ora}. The metric is given by
\begin{equation}
ds_5^2 ~=~{}  -\frac{1}{(Z_3 \cP)^{2/3}} \,(dt +  \kk)^2 \,+\,  (Z_3 \cP)^{1/3}\, ds_4^2 \,,
\label{eq:5D-metric}
\end{equation}
where
\be
\cP \;\equiv\; Z_1 Z_2 - Z_4^2\,.
\ee
The vector fields take the form, for $I=1,2,3,4$,
\be
A^{(I)} ~=~{} -\frac{C_{IJK} Z_J Z_K}{2\, Z_3 \cP} \left( dt + \kk \right) \,+\, B^{(I)} \,, 
\ee
where the constants $C_{IJK}$ are defined in Appendix \ref{app:BPS-equations} and where the magnetic components $B^{(I)}$ can be associated to four magnetic field strengths,
\be 
\Theta^{(I)} ~\equiv~{} d_{(4)} B^{(I)}.
\ee
where $d_{(4)}$ is the exterior derivative on the four-dimensional base space.

The functions $Z_I$ satisfy harmonic equations on the four-dimensional base with sources given by the wedge products of the magnetic two-form field strengths; more details can be found in Appendix \ref{app:BPS-equations}. These solutions can be uplifted to the $v$-independent class of the six-dimensional solutions that we will describe below \cite{Bena:2008wt} (for an alternative duality route see~\cite{Bena:2017geu}).

Let us start with some relatively well-known remarks on the asymptotics of the classes of solutions in which we are interested, before formulating our general \adstwo\ limit.
In the standard parameterization, a general class of solutions with the asymptotics of three-charge black holes in flat $\mathbb{R}^{1,4}$ is as follows. The base metric  $ds_4^2$ asymptotes to flat space, which we write in spherical polar coordinates:
\be \label{eq:4D-radial-asym}
ds^2_4 ~~\to~~ dr^2 + r^2 d\Omega_3^2 + \cdots ~~.
\ee
The functions $Z_\alpha$ ($\alpha=1,2,3$) asymptote to 1, and their subleading terms  give rise to the conserved charges $Q_\alpha$ of the solution:
\be
Z_\alpha ~~\to~~ 1 +  \frac{Q_\alpha}{r^2} + \cdots  \,, \qquad\qquad \alpha = 1,2,3 \,.
\ee
We consider solutions that have the same charges as the three-charge black hole, and hence have a $Z_4$ that decays at infinity faster than $1/r^2$.

We are interested in smooth horizonless ``bubbling solutions'', which have non-trivial topological structure that is controlled by another length-scale, $a$, and where there is a large hierarchy between the scale $a$ and the scales $Q_\alpha$,
\be \label{eq:a-ll-Q}
a^2 ~\ll~ Q_\alpha \,,
\ee
and where the $J_L$ angular momentum is inside the regime of parameters where the black hole horizon is macroscopic,
\be
\label{eq:jl-bound}
J_L^2 \;<\; Q_1 Q_2 Q_3 \,,
\ee
where we choose units and work at the appropriate locus in moduli space such that the dimensionful supergravity charges and the dimensionless quantized charges take the same values; thus in particular the five-dimensional gravitational coupling takes the value $G_5 = {\pi /4}$. 

In the regime
\be
a^2 ~\ll~ r^2 ~\ll ~ Q_\alpha \,,
\ee
the functions $Z_\alpha$ are approximately
\be
Z_\alpha ~~\simeq~~ \frac{Q_\alpha}{r^2} \,. 
\ee
For zero angular momenta, if one ``drops the constants'' in $Z_{\alpha}$, the asymptotics becomes \adstwo$\times$S$^3$. To illustrate this, we set $\kk$ to zero temporarily. We obtain
\be \label{eq:drop-const-Z}
Z_{\alpha} ~~\to~~  \frac{Q_{\alpha}}{r^2} + \cdots  \,, \qquad\qquad {\alpha} = 1,2,3 \,,
\ee
which, defining $Q \equiv (Q_1Q_2Q_3)^{1/3}$, yields:
\begin{equation}
ds_5^2 ~~\to~~  -\frac{r^4}{Q^{2}} \,dt^2 \,+\,  Q \frac{dr^2}{r^2}+ Q \, d\Omega_3^2 \,.
\label{eq:5D-metric-ads2}
\end{equation}
The $\adstwo\times$S$^3$ behaviour is manifest in coordinates $\hat{t} = 2t/\sqrt{Q}$, $\hat{r} = r^2/Q$, in which we have
\begin{equation}
ds_5^2 ~~\to~~ \frac{Q}{4} \left( - \hat{r}^2 \,d\hat{t}^2 \,+\,  \frac{d\hat{r}^2}{\hat{r}^2} \right) + Q \, d\Omega_3^2 \,.
\label{eq:5D-metric-ads2-2}
\end{equation}
This is the near-horizon geometry of the Strominger-Vafa black hole~\cite{Strominger:1996sh,Strominger:1998yg}. If one turns on a non-zero angular momentum charge, $J_L$,
one obtains the asymptotics of the near-horizon geometry of the BMPV black hole~\cite{Breckenridge:1996is},
\begin{equation}
\label{eq:NH-BMPV}
ds_5^2 ~~\to~~\frac{Q}{4}  \left[- \hat{r}^2 \! \left(d\hat{t}+\frac{J_L}{\hat{r}}\left(\sin^2\theta \, d\varphi_1 + \cos^2\theta \, d\varphi_2 \right) \right)^2 \!\! +  \frac{d\hat{r}^2}{\hat{r}^2}  \right]  \:+\: Q \,d\Omega_3^2 \;,
\end{equation}
where $\varphi_1$ and $\varphi_2$ are Cartan angles on the S$^3$. The BMPV black hole solution has no closed timelike curves and a macroscopic horizon when the angular momentum is inside the black hole regime, $J_L^2 < Q_1 Q_2 Q_3$.

\subsection{AdS$_2$ limit in five-dimensional supergravity}

We now describe a general limiting procedure to obtain an asymptotically \adstwo\ solution from an asymptotically flat bubbling solution satisfying the requirements \eq{eq:a-ll-Q} and \eq{eq:jl-bound}. 
We first extract the scale $a^2$ from the four-dimensional base metric,
\bea
ds_4^2 &=& a^2  \medtilde{ds}_4^2 \,,
\label{eq:base-rescale}
\eea
and we will be interested in taking a limit in which $a\to 0$ with $\medtilde{ds}_4^2$ and $Q_\alpha$ fixed. The scaling limit we shall derive is closely related to the families of ``scaling solutions'' mentioned above~\cite{Bena:2006kb,Denef:2002ru}, and is also closely related to other scaling limits considered previously in the literature for brane and black hole solutions (see for example \cite{Maldacena:1997re,Strominger:1998yg,Bardeen:1999px,Sen:2007qy,Kunduri:2013ana}).

We require that the scaling limit result in a non-singular solution. Let us derive the implications of this requirement for the dependence of the ansatz quantities on the parameter $a$. First, given the scaling of the base metric with $a$, \eq{eq:base-rescale}, in order to have a finite and non-trivial limit, as $a\to 0$ we must have
\be
Z_3 \cP ~\to~ \frac{\medtilde{Z}_3 \medtilde{\cP}}{a^6} 
\ee
where the product $\medtilde{Z}_3\medtilde{\cP}$ is finite and independent of $a$. This in turn implies that we must have
\be
\kk ~\to~ \frac{ \medtilde{\kk}}{a^2} \,, \qquad \qquad t ~=~ \frac{\tau}{a^2}
\label{eq:omega-rescale}
\ee
where $ \medtilde{\kk}$ is finite and independent of $a$, and where $\tau$ is held fixed as we take the limit.

Given a radial coordinate $r$ on the 4D base, that asymptotes to the radial coordinate in \eq{eq:4D-radial-asym}, we can thus define our AdS$_2$ limit by:
\be
r ~=~ a \:\! \tilde{r} \,, \qquad t ~=~ \frac{\tau}{a^2} \,\,; \qquad\qquad
\label{eq:ads2-limit}
a ~\to~ 0 \qquad \mathrm{with} ~~ \tilde{r} \,, \tau \,, Q_\alpha ~~ \mathrm{fixed} \,.
\ee

Examining the ansatz for the vector fields, one requires
\be
Z_I  ~\to~ \frac{\medtilde{Z}_I }{a^2}  \,, \qquad B^{(I)} ~\to~ \medtilde{B}^{(I)} \,,
\ee 
where $\medtilde{Z}_I$ and $\medtilde{B}^{(I)}$ are finite and independent of $a$. 
The above behavior of the ansatz quantities then ensures the finite limit
\bea
ds_5^2 &\to& {}  -\frac{1}{\phantom{\Bigl(}\!\!\!(\medtilde{Z}_3 \medtilde{\cP})^{2/3}} \,\big(d\tau +  \medtilde{\kk}\big)^2 \,+\, 
 (\medtilde{Z}_3 \medtilde{\cP})^{1/3}\, \medtilde{ds}_4^2 \,,
\label{eq:5D-metric-ads2-3} \\
A^{(I)} &\to& {} -\frac{C_{IJK} \medtilde{Z}_J \medtilde{Z}_K}{2\, \medtilde{Z}_3 \medtilde{\cP}\tildespace} \big( d\tau + \medtilde{\kk} \big) + \medtilde{B}^{(I)}\,. 
\label{eq:5D-vectors-ads2}
\eea
The limit $a\to 0$ has now been taken, and the solution is independent of $a$.

So far, we have treated the four-dimensional base metric completely generally. We next specialize by taking this base metric to have Gibbons-Hawking (GH) form,
\begin{equation}
ds_4^2 ~=~   V^{-1} \, (d \psi + A)^2   + V \;\! ds_3^2 \,,  \qquad ~~  \nabla^2 V ~=~ 0\,, \qquad  \ast_3 \:\! dV ~=~ dA \,
\label{GHmet2}
\end{equation}
where $\nabla^2$ is the Laplacian, $d$ is the exterior derivative, and $\ast_3$ is the Hodge star of the $\IR^3$ base.

If one assumes that all the ansatz quantities are independent of $\psi$, the general such solution can be written in terms of ten harmonic functions, $V, K^I, L_I, M$, on the $\IR^3$ base of this space (see Appendix \ref{app:BPS-equations} for details),
\be
Z_I = {1 \over 2}C_{IJK} {K^J K^K \over V} + L_I\,, \qquad \kk = (d\psi + A) \left( {1\over 6} C_{IJK} {K^I K^J K^K \over V^2} + {K^I L_I \over 2 V} + {M\over 2} \right)+ \varpi \,,~
\ee
where the angular momentum vector in the $\IR^3$ base of the GH space, $\varpi$, is determined by 
\be
\ast_3 \;\!\! d \varpi 
 ~=~ \frac{1}{2} \left( V \:\! d M - M d V + K^I d L_I - L_I d K^I\right) . 
\label{omega3D}
\ee
The absence of Dirac-Misner strings in $\varpi$ imposes certain constraints on the distances between the poles of the harmonic functions, which are known as ``bubble equations'' \cite{Bena:2005va,Denef:2000nb}.

In scaling solutions, the distances between the locations of the poles of the harmonic functions, as measured in the original three-dimensional base metric, tend to zero. As the points get closer and closer the throat becomes longer and longer, however the proper size of the various cycles supported by flux at the bottom of the throat stays finite in physical units \cite{Bena:2006kb,Bena:2007qc}. The \adstwo\ limit we are taking corresponds to zooming in on the bottom of the throat, while taking its length to infinity. This results in an asymptotically $\adstwo\times$S$^3$ solution.

Similarly to the above analysis, to obtain a finite and non-trivial limit we require
\be
V ~\to~ \frac{\medtilde{V}}{a^2} \,, \qquad ds_3^2 ~\to~ a^4 \;\! \medtilde{ds}_3^2  \;,
\ee
where $\medtilde{V}$, $\medtilde{ds}_3$ are finite and independent of $a$. 

Starting from the radial coordinate $\rho = r^2/4$ on the $\IR^3$ base of the GH space, one can define a rescaled radial coordinate analogously to \eq{eq:ads2-limit},
\be
\rho ~=~ a^2 \tilde\rho \,.
\label{eq:rholimit}
\ee
Then the coordinate $\tilde\rho$ is held fixed as we take $a \to 0$.

\newpage
Let us introduce a common notation for the ten harmonic functions,
\be
\mathcal{H}^{\Lambda} ~=~ \bigl\{V, K^I, L_I, M\bigr\} \,,
\ee
and parameterize a general multi-center solution as
\be
\mathcal{H}^{\Lambda}  ~=~ c^{\Lambda} + \sum_{A}\frac{q^\Lambda_A}{|\brho - \brho_A|} \,.
\ee
Then the scalings above correspond to the scaling of all the harmonic functions as
\be
\mathcal{H}^{\Lambda} \rightarrow \frac{\medtilde{\mathcal{H}}^{\Lambda}}{a^2} \,
\ee
where the $\mathcal{H}^{\Lambda}$ are finite and independent of $a$. 
In terms of the redefined coordinate three-vector on the base, $\tilde\brho$, the rescaled harmonic functions are
\be
\medtilde{\mathcal{H}}^{\Lambda} = a^2 \mathcal{H}^{\Lambda} = a^2c^{\Lambda} + \sum_{A}\frac{q^\Lambda_A}{|\tilde{\brho} -  {\tilde\brho}_A|} \,.
\ee
Thus the scaling limit, $a \to 0$, amounts to setting to zero the constants in all the harmonic functions, and therefore also the constants in $Z_I$; this connects to the discussion around \eq{eq:drop-const-Z}. Note that this does not change the charges of the solution, nor the number of magnetic flux quanta wrapping various two-cycles, nor the angular momentum along the GH fiber (corresponding to $J_L$ in five dimensions). This can be seen from the fact that these quantities are determined only by the coefficients of the poles in the harmonic functions, and are independent of the constants \cite{Bena:2005va,Berglund:2005vb}. 
However, the limit does set to zero the angular momentum $J_R$ along the $\IR^3 $ base of the GH space. This can be seen in two ways: from the scaling with $a$ of the $d \phi$ component of $\varpi$, and from the fact that in the absence of constants in the harmonic functions the right hand side of equation \eq{omega3D} decays too quickly to give rise to a finite angular momentum.

\subsection{AdS$_2$ limit in six-dimensional supergravity} 
\label{sec:AdS2-SS}

We now discuss the analogous procedure for a general set of BPS solutions to $\cN=(1,0)$ six-dimensional supergravity coupled to tensor multiplets. For concreteness we shall work with two tensor multiplets. In the D1-D5 system in Type IIB string theory compactified on $\cM=\,$T$^4$ or K3, and for configurations invariant on $\cM$, this system contains all the fields known to arise via worldsheet disk amplitudes for the backreaction of such bound states~\cite{Giusto:2011fy,*Black:2010uq}. It is straightforward to generalize the discussion to more general field content; indeed the general local form of BPS solutions in this theory, including more general matter, has recently been obtained~\cite{Cano:2018wnq,Lam:2018jln}. 

We consider solutions that have the asymptotics of a three-charge black string in $\IR^{4,1} \times$S$^1$. The asymptotic S$^1$ is coordinatized by $y$. Reducing along $y$, such a black string becomes the three-charge black hole in $\IR^{4,1}$, discussed in the previous subsection. 

Introducing the asymptotically null coordinate $v=t+y$, the six-dimensional metric is~\cite{Gutowski:2003rg,Giusto:2013rxa}
\begin{equation}
ds_6^2 ~=\, {}   -\frac{2}{\sqrt{\cP}} \, (dv+\beta) \big(du +  \kk - \tfrac{1}{2}\, Z_3\, (dv+\beta)\big) 
\,+\,  \sqrt{\cP} \, ds_4^2\,,
\label{sixmet}
\end{equation}
where 
\be
\cP \;=\;  Z_1 Z_2 - Z_4^2 \,.
\ee
There is some choice in how the 6D null coordinate $u$ is related to the 5D time coordinate $t$ upon reduction. We shall work in the parameterization in which $u=t$, which can be chosen as long as $Z_3$ is globally positive, as it will be in our solutions; for further discussion see~\cite{Bena:2017geu,Gutowski:2003rg}.

Given a form $\Phi$ with legs on the 4D base, depending on the coordinates of the 4D base and possibly on $v$, we define the operator $D$ via~\cite{Gutowski:2003rg}
\be
D \Phi \;\equiv\; d_{(4)}\Phi - \beta \wedge \partial_v \Phi\,.
\label{eq:D-op}
\ee
Our ansatz for the tensor fields $G^{(a)}$ ($a=1,2,4$) is then~\cite{Giusto:2013rxa,Bena:2017geu}
\be
G^{(a)}  \;=\;   d_{(4)} \left[ - \frac{1}{2}\,\frac{\eta^{ab} Z_b}{\cal P}\,(du + \kk ) \wedge (dv + \beta)\, \right] ~+~\coeff{1}{2} \;\! \eta^{ab} *_4 \! D Z_b  
~+~ \coeff{1}{2}\,  (dv+ \beta) \wedge \Theta^{(a)} \,,
\label{G-ans-cov}
\ee
where $\Theta^{(a)}$ are self-dual two-forms on the four-dimensional base, and $\ast_4$ is the Hodge star with respect to the four-dimensional base; more details, including the BPS equations, are given in Appendix \ref{app:6D}.
The dilaton and axion are given respectively by
\be
e^{2\varphi} \;=\; \frac{Z_1^2}{\cP} \,, \qquad \varsigma \;=\; \frac{Z_4}{Z_1} \,.
\ee

We again assume that we have a smooth horizonless solution with $J_L^2 < Q_1 Q_2 Q_3$ whose non-trivial structure is controlled by a length-scale, $a$, and that there is a large hierarchy $a^2 \ll Q_\alpha$  
\eq{eq:a-ll-Q}, such that the non-trivial structure of the solution is deep inside a throat that is S$^3$ fibered over extremal BTZ (which can be viewed as S$^1$ fibered over near-horizon BMPV), as in the solutions of~\cite{Bena:2016ypk,Bena:2017xbt}.

Following the same logic as in the previous subsection, we extract the scale $a^2$ from the 4D base metric as in \eq{eq:base-rescale}, and consider the limit $a\to 0$ with $\medtilde{ds}_4^2$ fixed.
Given a radial coordinate $r$ on the 4D base and writing $u=t$, we can define our AdS$_2$ limit by the following:
\be
r ~=~ a \:\! \tilde{r} \,, \qquad t ~=~ \frac{\tau}{a^2} \,\,; \qquad\qquad
\label{eq:ads2-limit-6D}
a ~\to~ 0 \qquad \mathrm{with} ~~ \tilde{r} \,, \tau \,, v  \,, Q_\alpha ~~ \mathrm{fixed} \,.
\ee
Since we hold $v$ fixed as $a \to 0$, we require $\beta$ to have a finite limit $\tilde{\beta}$ that is independent of $a$.
Given the scaling of the base metric with $a$, \eq{eq:base-rescale}, and examining the ansatz for the tensor fields \eq{G-ans-cov}, we see that to have a finite and non-trivial limit as $a\to 0$, we must have
\be \label{eq:Z-a-dep}
Z_I  ~\to~ \frac{\medtilde{Z}_I }{a^2}  \,, \qquad \kk ~\to~ \frac{ \medtilde{\kk}}{a^2} \,,
\qquad   \Theta^{(a)} ~\to~  \medtilde\Theta^{(a)}
\ee 
where $\medtilde{Z}_I$, $\medtilde{\kk}$, $\medtilde\Theta^{(a)}$ are finite and independent of $a$.
Note that the requirement in \eq{eq:Z-a-dep} on $\medtilde{Z}_I$, and in particular on $\medtilde{Z}_3$, relies crucially on the hierarchy $a^2 \ll Q_\alpha$ imposed in \eq{eq:a-ll-Q} and so, of course, does not hold for all solutions.
With these scalings, the ansatz becomes
\begin{equation}
ds_6^2 ~=~    -\frac{2}{\widetilde{\cP}^{1/2}} \, (dv+\tilde{\beta}) \left(d\tau +  \medtilde\kk - \tfrac{1}{2}\, \medtilde{Z}_3 \, (dv+\beta)\right) 
~+~  \widetilde{\cP}^{1/2} \, \medtilde{ds}_4^2 \,,
\label{sixmet-a2-2}
\end{equation}
\be
G^{(a)}  \;=\;   d_{(4)} \left[ - \frac{1}{2}\,\frac{\eta^{ab} \medtilde{Z}_b}{\medtilde{\cal P}\tildespace}\,(d\tau + \medtilde\kk ) \wedge (dv + \medtilde\beta)\, \right] ~+~\coeff{1}{2} \;\! \eta^{ab} \, \tilde{*}_4  D \medtilde{Z}_b  
~+~ \coeff{1}{2}\,  (dv+ \medtilde\beta) \wedge \medtilde\Theta^{(a)} \,,
\label{G-ans-cov-2}
\ee
\be
e^{2\varphi} \;=\; \frac{\medtilde{Z}_1^2}{\medtilde\cP\tildespace} \,, \qquad \varsigma \;=\; \frac{\medtilde{Z}_4}{\medtilde{Z}_1\tildespace} \,,
\ee
where $\tilde{*}_4$ is the Hodge star with respect to $\medtilde{ds}_4$. At this point, the $a\to 0$ limit has been taken, and the ansatz quantities no longer depend on $a$.

\subsection{Extremal BTZ throat inside AdS$_3$}

We now consider the interesting regime of parameters in which one of the charges (say $Q_3$) is much smaller than the others, but still larger than the scale $a$ controlling the non-trivial topological structure of the solution,
\be \label{eq:QI-hierarchy}
a^2 ~\ll~ Q_3 ~\ll~ \{ Q_{1} , Q_2 \} \,.
\ee
In this regime, one has an approximate extremal BTZ throat~\cite{Banados:1992wn} inside an approximate \adsthree\ throat. The extremal BTZ throat can be viewed as S$^1$ fibered over \adstwo.

In the regime of parameters~\eq{eq:QI-hierarchy}, let us review the relation of the above procedure to ``dropping the constants'' progressively. 
Flat $\IR^{4,1} \times S^1$ asymptotics corresponds to $Z_1$, $Z_2$, $Z_3$ having the large-$r$ asymptotics
\be
Z_1 ~\sim~ 1 + \frac{Q_1}{r^2} + \cdots \,, \qquad
Z_2 ~\sim~ 1 + \frac{Q_2}{r^2} + \cdots \,,\qquad
Z_3 ~\sim~ 1 + \frac{Q_3}{r^2} + \cdots \,.
\ee
Taking a scaling limit to an asymptotically AdS$_3 \times$S$^3$ solution involves ``dropping the constants'' in $Z_1$ and $Z_2$ only,
\be
Z_1 ~\sim~  \frac{Q_1}{r^2} + \cdots \,, \qquad
Z_2 ~\sim~  \frac{Q_2}{r^2} + \cdots \,,\qquad
Z_3 ~\sim~ 1 + \frac{Q_3}{r^2} + \cdots \,.
\ee
Taking a further limit to an asymptotically AdS$_2 \times$S$^1\times$S$^3$ solution corresponds to  also ``dropping the constant'' in $Z_3$, such that 
\be
Z_I ~\sim~  \frac{Q_I}{r^2} + \cdots \,, \qquad\quad I = 1,2,3 \,.
\ee

Let us compare the last step with the above limit. We have
\bea
Z_3 &\simeq& 1 + \frac{Q_3}{a^2 \tilde{r}  ^2}  + \cdots 
\eea
and so we see that as $a\to 0$ with fixed $Q_3$, the second term dominates the first, and so the constant is indeed ``dropped'' in the limit  $a\to 0$. We note also that the term that survives has the required $a$-dependence \eq{eq:Z-a-dep}.

Let us emphasize that our requirements that $Q_\alpha \gg a^2$ and $J_L^2 < Q_1 Q_2 Q_3$ do not hold for all BPS solutions: examples include of course two-charge BPS solutions with $Q_3=0$~\cite{Lunin:2001fv,*Lunin:2001jy,Emparan:2001ux,Lunin:2002iz,Kanitscheider:2007wq}, and also certain atypical three-charge BPS solutions that have $J_L$ outside the black hole regime (see for example \cite{Giusto:2004id,*Giusto:2004ip,Jejjala:2005yu,Giusto:2012yz}). However, typical BPS black hole microstates will indeed satisfy both $Q_\alpha \gg a^2$ and $J_L^2 < Q_1 Q_2 Q_3$, and the above discussion will thus apply.

\section{Asymptotically AdS$_2$ multi-center solutions}
\label{sec3}

For five-dimensional solutions that can be expressed in terms of harmonic functions, in the previous section we saw that our general \adstwo\ limit amounts to setting to zero all the constants in the harmonic functions. In this section we will exploit this fact together with the systematic procedure developed in \cite{Heidmann:2017cxt} to construct a novel family of smooth horizonless solutions
that have four Gibbons-Hawking centers and that are free of closed timelike curves (CTCs). All the solutions constructed in this section have the asymptotics of the near-horizon BMPV solution; we will continue to frequently use the convenient abuse of notation of describing these solutions as asymptotically \adstwo.

We work in $\cN=1$ five-dimensional STU (or $U(1)^3$) supergravity, which can be embedded in the $U(1)^4$ model described in the previous section and Appendix \ref{app:BPS-equations} by switching off the fourth $U(1)$:
\be 
K_4 ~=~ 0, \qquad L_4 ~=~ 0, \qquad Z_4 ~=~ 0 \qquad \Rightarrow \quad \cP ~=~ Z_1 Z_2.
\ee

To obtain asymptotically AdS$_2$ solutions we will use the method introduced in  \cite{Heidmann:2017cxt}, adapted to the absence of constant terms in the harmonic functions as explained in the previous section. We restrict to solutions where all the points lie in a plane, and denote by  $\brho_A$  the position vector of center $A$ in the three-dimensional base space of the solution. The distance from the origin (center 0) is given by its absolute value, $\rho_A$, and its angle (with respect to an arbitrarily chosen axis in the plane) is $\phi_A$. We order the positions of the centers as 
\begin{equation}
\rho_{1} \:\geq\: \rho_{2}  \:\geq\: \rho_{3} \:\geq\: \rho_0 \:=\: 0.
\end{equation}
The full solution is determined by eight harmonic functions \cite{Gutowski:2004yv,*Gauntlett:2004qy,*Bena:2004de,Bena:2005ni}, 
\begin{equation}
\label{Z&kexpression}
\begin{split}
    & V\left(\brho\right) \:=\: q_{\infty} \:+\: \sum_{A=0}^{3} \frac{q_A}{|\brho -\brho_A|}\,, \qquad M\left(\brho\right) \:=\: m_{\infty} \:+\: \sum_{A=0}^{3} \frac{m_A}{|\brho -\brho_A|} \,, \\
    & K^I\left(\brho\right) \:=\: k_{\infty}^I \:+\: \sum_{A=0}^{3} \frac{k_A^I}{|\brho -\brho_A|}\,, \qquad L_I\left(\brho\right) \:=\: l_{\infty}^I \:+\: \sum_{A=0}^{3} \frac{l_A^I}{|\brho -\brho_A|}\,,
\end{split}
\end{equation}
where $\brho$ is the position vector in the three-dimensional base space of the solution. We recall that the radial coordinate $\rho=|\brho|$ is related to the radial coordinate $r$ of the four-dimensional base as $\rho = {r^2 / 4}$.

\subsection{Construction procedure}

If one chooses the eight harmonic functions \eqref{Z&kexpression} at random, one will almost certainly obtain a
solution that is singular and has closed timelike curves (CTCs). To systematically construct smooth, CTC-free scaling solutions with four Gibbons-Hawking centers we  perform generalized spectral flows and gauge transformations on a solution with three supertube centers and a Gibbons-Hawking (GH) center, following \cite{Heidmann:2017cxt}. Generalized spectral flows are linear transformations on the harmonic functions \cite{Bena:2008wt} and hence transform one asymptotically \adstwo\ solution (with no constants in the harmonic functions) into another. Solving the bubble equations and imposing absence of CTCs is much easier for two-charge supertubes than for GH centers. Note that generalized spectral flows and gauge transformations preserve the entropy of the corresponding black hole, as well as the bubble equations. 
 
 The solutions constructed in \cite{Heidmann:2017cxt} had $\IR^{4,1}$ asymptotics, and the constant terms in their harmonic functions were $\left(q_\infty , k^1_\infty , k^2_\infty , k^3_\infty ; l^1_\infty ,  l^2_\infty , l^3_\infty , m_\infty \right) = \left(0,0,0,0;1,1,1,m_\infty \right)$.   We give here the two remaining steps to obtain solutions with no constant terms (and hence \adstwo\ asymptotics):

\begin{itemize}
\item Let us first detail how to set to zero the constant terms of the harmonic functions. We do this by hand, without changing the values of the charges, and ensuring that smoothness is preserved. The impact of a change in the constant terms is visible mostly on the bubble equations. One may therefore naively expect that in the scaling limit a change of constants can be compensated by an infinitesimal change in the distances. However, this intuition is not correct for axisymmetric configurations, as will be explained in upcoming work \cite{Heidmann:2018toappear}. To remove all the constant terms one needs to break the axisymmetry,  by allowing at least one of the angles to vary as one approaches the scaling limit. 

\item To obtain a solution with quantized charges and fluxes, one can scale up the harmonic functions using the transformation
\begin{equation}
M \:\rightarrow\: g^3 M, \qquad L^{I} \:\rightarrow\: g^2 L^{I}, \qquad 
  K^{I} \:\rightarrow\: g\, K^{I}, \qquad V \:\rightarrow\: V,  \qquad~~~ g\in\mathbb{R} \,.
\label{overallmultiplication}
\end{equation}
When there are no constant terms in the harmonic functions, this leaves the bubble equations invariant. Thus, if one chooses $g$ to be the lowest common multiple of the denominators of  $k^{I}$ one obtains a smooth horizonless asymptotically \adstwo\ solution with quantized charges and fluxes.
\end{itemize}

\subsection{An explicit example}

In this section, we construct an explicit scaling BPS solution with four Gibbons-Hawking centers which is asymptotically AdS$_2$, following the procedure outlined above. We assume that we have taken the limit $a \rightarrow 0$ defined in Eq.\;\eq{eq:ads2-limit}, \eq{eq:rholimit} and we drop the tildes in Eq.\;\eq{eq:rholimit} for readability. We choose the inter-center distances with no hierarchy in scale between them,
\begin{equation}
\frac{\rho_1 - \rho_2}{\rho_3} \:\approx\: 1 \,, \qquad\quad
\frac{\rho_2 - \rho_3}{\rho_3} \:\approx\: 1 \,.
\label{distanceratios}
\end{equation}
We start with a particular configuration of three supertubes of different kinds and a Gibbons-Hawking center and we apply all the steps of the procedure. We end up with the following solution:
\begin{equation}
\begin{split}
& V \:=\: \frac{1}{|\brho\, |} \:-\: \frac{2}{|\brho -\brho_1|} \:+\: \frac{1}{|\brho -\brho_2|} \:+\: \frac{1}{|\brho -\brho_3|} \\
& K^1 \:=\:  -\frac{21}{|\brho\, |} \:-\: \frac{138}{|\brho -\brho_1|} \:+\: \frac{56}{|\brho -\brho_2|} \:-\: \frac{27}{|\brho -\brho_3|}  \\
& K^2 \:=\:  \frac{47}{|\brho\,|} \:-\: \frac{4210}{|\brho -\brho_1|} \:+\: \frac{1055}{|\brho -\brho_2|} \:+\: \frac{1628}{|\brho -\brho_3|} \\
& K^3 \:=\:  \frac{86}{|\brho\,|} \:-\: \frac{20}{|\brho -\brho_1|} \:-\: \frac{196}{|\brho -\brho_2|} \:-\: \frac{79}{|\brho -\brho_3|} \\
& L^1 \:=\:  -\frac{4042}{|\brho\,|} \:+\: \frac{42100}{|\brho -\brho_1|} \:+\: \frac{206780}{|\brho -\brho_2|} \:+\: \frac{128612}{|\brho -\brho_3|}  \\
& L^2 \:=\: \frac{1806}{|\brho\,|} \:+\: \frac{1380}{|\brho -\brho_1|} \:+\: \frac{10976}{|\brho -\brho_2|} \:-\: \frac{2133}{|\brho -\brho_3|}  \\
& L^3 \:=\:  \frac{987}{|\brho\,|} \:+\: \frac{290490}{|\brho-\brho_1|} \:-\: \frac{59080}{|\brho -\brho_2|} \:+\: \frac{43956}{|\brho -\brho_3|}  \\
& M \:=\: -\frac{84882}{|\brho\,|} \:-\: \frac{2904900}{|\brho -\brho_1|} \:-\: \frac{11579680}{|\brho -\brho_2|} \:+\: \frac{3472524}{|\brho -\brho_3|} \,.
\label{NewSol}
\end{split}
\end{equation}
The bubble equations are solvable, and the resulting positions of the centers are given by
\begin{alignat}{2}
& \frac{\rho_1 - \rho_2}{\rho_3} \:=\: 0.87126\ldots \,, \qquad && \frac{\rho_2 - \rho_3}{\rho_3} \:=\: 0.81999\ldots \, ,\\
& \phi_1 \:=\: \phi_3 \:=\: 0 \,, \qquad && \phi_2 \:=\: 12.565\ldots\degree ~.
\end{alignat}
The three charges of the solution obtained from the asymptotics of the $Z_I$:
\begin{equation}
Q_1 \:=\: 682770 \,, \qquad Q_2 \:=\: 39199 \,, \qquad Q_3 \:=\: 468753 \,.
\end{equation}
For the angular momenta, one has
\begin{equation}
\kk \:\sim\: \left[ \frac{J_L + J_R \, \cos \theta}{\rho} + \mathcal{O}\left(\frac{1}{\rho^2}\right)\right](d\psi + A) \,+\, \mathcal{O}\left(\frac{1}{\rho^2}\right) d\phi,
\end{equation}
with 
\begin{equation}
J_L \;=\; -\frac{215629335}{2}  \;\simeq\;  0.96\,J_{\rm max}\,, \qquad~~~ J_R \;=\; 0 \,,
\end{equation}
where $J_{\rm max}\equiv\sqrt{Q_1 Q_2 Q_3}$ is the cosmic censorship bound of the black hole with these charges.

Note that $J_R$ is strictly equal to 0, which does not happen in general for flat or $AdS_3$ asymptotics. Indeed, for generic scaling multicenter solutions, $J_R$ is non-zero and inversely proportional to the length of the AdS$_2$ throat. In order for such solutions to look more and more like a black hole, they must be taken to the scaling limit, in which the size of the throat becomes infinitely long. Our procedure to obtain asymptotically AdS$_2$ solutions can therefore be thought of as approaching the scaling limit while lowering the energy of our observers. In this limit the bubble equations and the physical sizes of the bubbles become invariant under uniform rescalings of the GH radial coordinate.

Finally, the Bekenstein-Hawking entropy of the corresponding black hole is
\begin{equation}
S \:=\: 3.036\ldots \times 10^7,
\end{equation}
and far from all the bubbles, when $|\brho -\brho_A| \gg |\brho_A-\brho_B|$, the five-dimensional metric becomes approximately
\begin{equation}
ds^2_{5} \sim (Q_1 Q_2 Q_3)^{\frac{1}{3}}  \left[-\hat{\rho}^2 \! \left(d\hat{t}+\frac{J_L}{\hat{\rho}}(d\psi+\cos\theta d\phi)\right)^2 \!\! +  \frac{d\hat{\rho}^2}{\hat{\rho}^2}  \right.  \:+\:  \left. \vphantom{\left(\frac{A}{A}\right)}d\theta^2 + \sin^2 \theta  d\phi^2 + (d\psi+\cos\theta d\phi)^2\] 
\end{equation}
where $\hat{\rho} = (Q_1 Q_2 Q_3)^{-1} \rho$ and $\hat{t} = \sqrt{Q_1 Q_2 Q_3} \,t$. We recognize this as the five-dimensional near-horizon BMPV solution \eq{eq:NH-BMPV}, which from a two-dimensional perspective is AdS$_2$ with matter fields. 
We have thus constructed an explicit example of a microstate geometry with near-horizon BMPV asymptotics and four Gibbons-Hawking centers.

\section{Asymptotically AdS$_2$ superstrata}
\label{sec4}

In this section we apply the six-dimensional AdS$_2$ limit defined in Section \ref{sec:general-limit} to construct families of explicit supergravity solutions with near-horizon-BTZ$\times$S$^3$ asymptotics, which as described above can be written as S$^1$ fibered over the near-horizon BMPV solution. We will exhibit an explicit family of examples that are asymptotically S$^1$ fibered over \adstwo$\times$S$^3$.

\subsection{Superstrata with a flat base metric}

The procedure discussed in Section \ref{sec:AdS2-SS} is general, and can be applied to all solutions which have the hierarchy of charges (\ref{eq:a-ll-Q}). For concreteness we will work with the family of superstratum solutions for which the 4D base is flat $\IR^4$, written as 
\bea
ds_4^2 = \Sigma\, \bigg( \frac{dr^2}{r^2 + a^2} \,+\, d \theta^2 \bigg)  \,+\, (r^2 + a^2) \sin^2 \theta \, d\varphi_1^2 \,+\, r^2  \cos^2 \theta \, d\varphi_2^2 \,,
\eea
where
\be
\Sigma ~=~ r^2 + a^2 \cos^2\theta \,,
\ee
and where the one-form $\beta$ is
\bea
\beta &=& \frac{R_y a^2 }{\Sigma} \;\! ( \sin^2 \theta \, d\varphi_1-   \cos^2 \theta \, d\varphi_2 )\,.
\eea
Then we see that in the above limit \eq{eq:ads2-limit-6D},
\bea
\beta ~\to~\medtilde{\beta} &=& \frac{R_y}{\tilde{r}^2 + \cos^2\theta } \;\! ( \sin^2 \theta \, d\varphi_1-   \cos^2 \theta \, d\varphi_2 )\,,
\eea
which is of order $a^0$, as required. For the 4D base, we obtain
\bea
\medtilde{ds}_4^2 &=& (\tilde{r}^2 + \cos^2\theta )\, \bigg( \frac{d\tilde{r}^2}{\tilde{r}^2 + 1} \,+\, d \theta^2 \bigg)  \,+\, (\tilde{r}^2 + 1) \sin^2 \theta \, d\varphi_1^2 \,+\, \tilde{r}^2  \cos^2 \theta \, d\varphi_2^2 \,.
\eea
The family of solutions with a flat base is a large one, and includes in particular the solutions of~\cite{Balasubramanian:2000rt,*Maldacena:2000dr,Bena:2015bea,Bena:2016agb,Bena:2016ypk,Bena:2017xbt}.

\subsection{Single-mode superstrata}
\label{subsec:singlemodeS}

We now focus attention further to the single-mode superstratum solutions recently constructed in~\cite{Bena:2016ypk,Bena:2017xbt}. This is a sub-family of superstrata that is relatively easy to study, while having many interesting features. The solutions we consider are labelled by three positive integers 
\be
(k,m,n) \,, \qquad\quad k \ge 1 \,, \quad 0 \le m \le k \,, \quad n \ge 0 \,,
\ee
and a continuous parameter $a/b$, where the length scales $a$ and $b$ are constrained to obey:
\be
a^2 + \frac{b^2}{2} ~=~ \frac{Q_1 Q_2}{R_y^2} \,.
\ee
For each allowed value of the triple $(k,m,n)$, there is an asymptotically $\mathbb{R}^{4,1}\times$S$^1$ solution as well as an asymptotically AdS$_3 \times $S$^3$ solution; in the asymptotically AdS$_3$ solutions, the metric preserves four isometries, however three of these isometries are broken by the explicit dependence of the matter fields on the phase~\cite{Bena:2016ypk}
\be
\hat{v}_{k,m,n} ~\equiv~ (m+n) \frac{v}{R_y} + (k-m)\varphi_1 - m\varphi_2 \,.
\ee
In the full asymptotically flat solutions, the metric also depends on the above phase~\cite{Bena:2017xbt}. We shall apply our limit to the asymptotically AdS$_3 \times $S$^3$ solutions. 

We will soon focus for our explicit presentation on solutions that have $(k,m,n)=(1,0,n)$, however for the moment, and where it is illuminating, we shall keep $(k,m,n)$ general, to illustrate the generality of the procedure. 
The momentum charge along the asymptotic AdS$_3$ circle coordinatized by $y$ is given by
\bea
Q_3 &=& \frac{b^2}{2}\frac{m+n}{k} \,.
\label{eq:Q3-val-SS}
\eea
Following the general discussion in Section \ref{sec:AdS2-SS},
in the regime
\bea \label{eq:ads2-regime-SS}
a^2 ~\ll ~ Q_3 ~\ll ~ \{ Q_1 \,, Q_2 \} \,,
\eea
we have a BTZ-like near-horizon throat inside AdS$_3 \times $S$^3$. 

To write the solutions explicitly, we introduce the notation
\bea
\Delta_{k,m,n} &\equiv&
 \left(\frac{a}{\sqrt{r^2+a^2}}\right)^k
 \left(\frac{r}{\sqrt{r^2+a^2}}\right)^n 
 \cos^{m}\theta \, \sin^{k-m}\theta \,,  \label{Delta_v_kmn_def} \\
\vartheta_{k,m,n}
 &\equiv & -\Delta_{k,m,n}
 \biggl[\left((m+n)r\sin\theta+n\left({m\over k}-1\right){\Sigma\over r \sin\theta}  \right)\Omega^{(1)}\sin\vh_{k,m,n}\notag\\
 &&\hspace{15ex}
 +\left(m\left({n\over k}+1\right)\Omega^{(2)}
 +n\left({m\over k}-1\right)\, \Omega^{(3)} \right)\cos\vh_{k,m,n}\biggr] \,,
\eea 
where $\Omega^{(i)}$ ($i=1,2,3$) are a basis of self-dual 2-forms on
$\mathbb{R}^4$:
\begin{equation}\label{selfdualbasis}
\begin{aligned}
\Omega^{(1)} &\equiv \frac{dr\wedge d\theta}{(r^2+a^2)\cos\theta} + \frac{r\sin\theta}{\Sigma} d\varphi_1\wedge d\varphi_2 \,,\\
\Omega^{(2)} &\equiv  \frac{r}{r^2+a^2} dr\wedge d\varphi_2 + \tan\theta\, d\theta\wedge d\varphi_1\,,\\
 \Omega^{(3)} &\equiv \frac{dr\wedge d\varphi_1}{r} - \cot\theta\, d\theta\wedge d\varphi_2\,.
\end{aligned}
\end{equation}

In order to more easily connect to the discussion of ``dropping the 1'' in the function $Z_3$, we will write these solutions in the following parameterization (see~\cite{Bena:2017geu} for a discussion of how this ``$u=t$''  parameterization is related to that presented in~\cite{Bena:2016ypk,Bena:2017xbt}). The solution to the first layer is given by
\begin{equation}
\begin{aligned}
Z_1 ~=~ & \frac{Q_1}{\Sigma} + \frac{b_1 R_y^2 }{2 Q_2} \,
\frac{\Delta_{2k,2m,2n}}{\Sigma} \cos \vh_{2k,2m,2n} \,,   \qquad Z_2 ~=~ \frac{Q_2}{\Sigma} \,,  \\
Z_4~=~   & R_y  \, b_4\, \frac{ \Delta_{k,m,n} }{\Sigma} \cos \vh_{k,m,n}  \,,
\end{aligned}
 \label{eq:ZIAdSsinglemode} 
\end{equation}
with
\begin{equation}
\label{eq:ThetaIAdSsinglemode}
 \Theta_1 =0\,,\qquad
 \Theta_2 = \frac{b_1 R_y}{2 Q_2}\, \vartheta_{2k,2m,2n}\,,\qquad
 \Theta_4 = b_4\, \vartheta_{k,m,n}\,,
\end{equation}
and where 
\begin{equation} 
b_1 ~=~b_4^2\,.
\label{AdScoiff}
\end{equation} 

The solution to the second layer for general $(k,m,n)$ was found in \cite{Bena:2016ypk}. For ease of presentation, at this point we will specialize to the sub-family $(k,m,n)=(1,0,n)$, as this will suffice for an explicit family of examples. It is straightforward to generalize the following discussion to the general $(k,m,n)$ family. The solution to the second layer for the $(1,0,n)$ family is~\cite{Bena:2016ypk}:
\bea
Z_3 &=& 1 + \frac{b_4^2}{2 a^2} \left( 1 - \frac{r^{2n}}{(r^2+a^2)^n} \right)    \,,  \cr
\kk &=& \kk_0 + \frac{b_4^2 R_y}{2 \;\! \Sigma} \left( 1 - \frac{r^{2n}}{(r^2+a^2)^n} \right) \sin^2 \theta \, d\varphi_1  \,,
\eea
where
\be
\kk_0 ~=~ \frac{a^2 R_y}{\Sigma} \sin^2 \theta \, d\varphi_1 \,.
\ee

The solution in which $b_4=0$ is simply the round circular supertube solution of~\cite{Balasubramanian:2000rt,*Maldacena:2000dr}. Thus these superstratum solutions can be regarded as a family of smooth finite deformations of a supertube. The regime in which the AdS$_2$ throat exists, \eq{eq:ads2-regime-SS}, is that in which the deformation dominates over the original supertube component of the solution. 

We now take the AdS$_2$ limit of this family of solutions. As before, to write the resulting expressions, we will drop the tilde on the radial coordinates in \eq{eq:ads2-limit-6D}. 
We will continue to write the general $(k,m,n)$ expressions for the first-layer data, specializing to $(1,0,n)$ for the second layer.

The phase $\hat{v}_{k,m,n}$ is invariant; we define the quantities
\bea
\medtilde{\Sigma} &\equiv& \tilde{r}^2 + \cos^2\theta \,, \cr
\medtilde{\Delta}_{k,m,n}  &\equiv&
 \left(\frac{1}{\sqrt{\tilde{r}^2+1}}\right)^k
 \left(\frac{\tilde{r}}{\sqrt{\tilde{r}^2+1}}\right)^n 
 \cos^{m}\theta \, \sin^{k-m}\theta \,,  \label{Delta_v_kmn_def-AdS2} \\
\medtilde{\vartheta}_{k,m,n}
 &\equiv & -\medtilde{\Delta}_{k,m,n}
 \biggl[\left((m+n)\tilde{r}\sin\theta+n\left({m\over k}-1\right){\medtilde{\Sigma} \over \tilde{r} \;\!\sin\theta}  \right)\medtilde\Omega^{(1)}\sin\vh_{k,m,n}\notag\\
 &&\hspace{15ex}
 +\left(m\left({n\over k}+1\right)\medtilde\Omega^{(2)}
 +n\left({m\over k}-1\right)\, \medtilde\Omega^{(3)} \right)\cos\vh_{k,m,n}\biggr] , \nonumber
\eea 
where $\medtilde\Omega^{(i)}$ ($i=1,2,3$) are given by:
\begin{equation}\label{selfdualbasis-2}
\begin{aligned}
\medtilde\Omega^{(1)} &\equiv \frac{d\tilde{r}\wedge d\theta}{(\tilde{r}^2+1)\cos\theta} +
 \frac{\tilde{r}\sin\theta}{\medtilde{\Sigma}\phantom{{\medhat{\Sigma}}^{(9)}}\hspace{-6.8mm}} d\varphi_1\wedge d\varphi_2\,,\\
\medtilde\Omega^{(2)} &\equiv  \frac{\tilde{r}}{\tilde{r}^2+1} d\tilde{r}\wedge d\varphi_2 + \tan\theta\, d\theta\wedge d\varphi_1\,,\\
 \medtilde\Omega^{(3)} &\equiv \frac{d\tilde{r}\wedge d\varphi_1}{\tilde{r}} - \cot\theta\, d\theta\wedge d\varphi_2\,.
\end{aligned}
\end{equation}
and we see that in the AdS$_2$ limit,
\bea
\Omega^{(1)} ~\to~ a \;\! \medtilde\Omega^{(1)} \,, \qquad \Omega^{(2)} ~\to~ \medtilde\Omega^{(2)} \,, \qquad \Omega^{(3)} ~\to~ \medtilde\Omega^{(3)} \,.
\eea
The tilded quantities defined in \eq{eq:Z-a-dep} for the first-layer data, become
\begin{equation}
\begin{aligned}
\medtilde{Z}_1 ~=~ & \frac{Q_1}{\medtilde{\Sigma}\phantom{{\medhat{\Sigma}}^{(9)}}\hspace{-6.8mm}} + \frac{b_1 R_y^2 }{2 Q_2} \,
\frac{\medtilde{\Delta}_{2k,2m,2n}}{\medtilde{\Sigma}\phantom{{\medhat{\Sigma}}^{(9)}}\hspace{-6.8mm}} \cos \vh_{2k,2m,2n} \,, 
\qquad \medtilde{Z}_2 ~=~ \frac{Q_2}{\medtilde{\Sigma}\phantom{{\medhat{\Sigma}}^{(9)}}\hspace{-6.8mm}} \,,  \\
\medtilde{Z}_4~=~   & R_y  \, b_4\, \frac{ \medtilde{\Delta}_{k,m,n} }{\medtilde{\Sigma}\phantom{{\medhat{\Sigma}}^{(9)}}\hspace{-6.8mm}} \cos \vh_{k,m,n}  \,,
\end{aligned}
 \label{eq:ZIAdSsinglemode-AdS2} 
\end{equation}
and
\begin{equation}
\label{eq:ThetaIAdSsinglemode-AdS2}
\medtilde\Theta_1 =0\,,\qquad
\medtilde\Theta_2 = \frac{b_1 R_y}{2 Q_2}\, \medtilde\vartheta_{2k,2m,2n}\,,\qquad
 \medtilde\Theta_4 = b_4\, \medtilde\vartheta_{k,m,n}\,.
\end{equation}

To write the second-layer quantities, we specialize again to the $(1,0,n)$ solution, obtaining:
\bea
\medtilde{Z}_3 &=& \frac{b_4^2}{2} \left( 1 - \frac{\tilde{r}^{2n}}{(\tilde{r}^2+1)^n} \right)    \,,  \cr
\medtilde{\kk} &=& \frac{b_4^2}{2}\frac{R_y}{\medtilde{\Sigma}\phantom{{\medhat{\Sigma}}^{(9)}}\hspace{-6.8mm}} \left( 1 - \frac{\tilde{r}^{2n}}{(\tilde{r}^2+1)^n} \right) \sin^2 \theta \, d\varphi_1  \,.
\eea
Smoothness of the $(1,0,n)$ solution imposes the relation
\be
b ~=~ b_4 \,.
\ee
The general $(k,m,n)$ smoothness relation between $b$ and $b_4$ can be found in~\cite{Bena:2016ypk,Bena:2017xbt}.

We observe that our \adstwo\ limit  can be described as ``dropping the $1$'' in $Z_3$, and also dropping the associated supertube component of $\kk$, namely $\kk_0$.
It is not hard to see that the second layer of the BPS equations (given in Appendix \ref{app:6D}) are still satisfied, precisely because $\kk_0$ balances the ``$1$'' in $Z_3$.

To write the full metric, we define the following shorthands:
\bea
\medtilde{F}_0(\tilde{r}) &\equiv& 1 - \frac{\tilde{r}^{2n}}{(\tilde{r}^2+1)^n} \,, \qquad \medtilde{F}_1(\tilde{r}) ~\equiv~ 1 - \frac{\tilde{r}^{2n}}{(\tilde{r}^2+1)^{n+1}}\,,\cr
\medtilde{\Lambda} &\equiv& \frac{\medtilde{\Sigma}\sqrtpadstwo}{\sqrt{Q_1 Q_2}} ~=~ \sqrt{1-\frac{\tilde{r}^{2n}}{(\tilde{r}^2+1)^{n+1}}\sin^2\theta} \,.
\label{eq:F0tilde}
\eea
We will write the full metric in two ways. The first is more convenient for displaying the AdS$_2$ asymptotics, the second is more convenient to see the smoothness in the cap. In the first form of the metric, the square is first completed on the $dv$ terms, as appropriate for a reduction from 6D to 5D on the $v$ fiber,
\bea
ds_6^2 &\!=\!& Q_1 Q_2 \frac{\medtilde{F}_0(\tilde{r})}{\sqrtpadstwo}\left( \frac{dv}{R_y} - \frac{1}{\medtilde{F}_0(\tilde{r}) \tildespace} \frac{2 \, d\tau}{b^2 R_y} - \frac{\cos^2\theta}{\sigmaadstwodenom} d\varphi_2 \right)^2
-\frac{2}{b^2\sqrt{\medtilde{\cP}\tildespace}}\left( \frac{1}{\medtilde{F}_0(\tilde{r})\tildespace} + \frac{\sin^2\theta}{\sigmaadstwodenom}  \right)d\tau^2\cr
&&{} + \sqrt{Q_1 Q_2} \; \medtilde{\Lambda} \left( \frac{d\tilde{r}^2}{\tilde{r}^2+1}+ d\theta^2 + \frac{\rt^2 \cos^2\theta}{\sigmaadstwodenom} d\varphi_2^2 \right)
+\frac{\sqrt{Q_1 Q_2}}{\medtilde{\Lambda} \tildespace}\sin^2\theta \left( d\varphi_1  - \frac{2 \, d\tau}{b^2 R_y}   \right)^2 \!\!.
\qquad~~~
\label{eq:6dmetric1}
\eea
At large $\rt$, this metric becomes that of the ``very-near-horizon'' limit of the 6D non-rotating black string, in the form in which the $v$ direction is fibered over the very-near-horizon limit of the five-dimensional non-rotating supersymmetric (Strominger-Vafa) black hole~\cite{Strominger:1998yg} (c.f.~Eq.\;\eq{eq:5D-metric-ads2}),
\bea
ds_6^2 &\!\!=\!\!& \frac{Q_3}{\sqrt{Q_1 Q_2}}\left( dv - \frac{\tilde{r}^2}{Q_3} d\tau \right)^2
-\frac{\tilde{r}^4 d\tau^2}{Q_3 \sqrt{Q_1 Q_2}} 
+ \sqrt{Q_1 Q_2}  \left( \frac{d\tilde{r}^2}{\tilde{r}^2}+ d\theta^2 + \sin^2\theta  d\varphi_1^2 + \cos^2\theta d\varphi_2^2  \right) 
\cr
&\!\!=\!\!& - 2 \frac{\tilde{r}^2 dv \:\! d\tau}{\sqrt{Q_1 Q_2}} + \frac{Q_3}{\sqrt{Q_1 Q_2}}  dv^2
+ \sqrt{Q_1 Q_2} \left( \frac{d\tilde{r}^2}{\tilde{r}^2}+ d\theta^2 + \sin^2\theta  d\varphi_1^2 + \cos^2\theta d\varphi_2^2  \right) . \phantom{\Bigg)}
\label{eq:6dmetric1-asym}
\eea

We next write the metric in a second form in the squares are completed first on the S$^3$ directions, demonstrating the smooth shrinking of the remaining directions in the cap. The metric in this form is given by
\bea
ds_6^2 &=& - \frac{2}{b^2}\frac{\medtilde{\Lambda}}{\sqrt{Q_1 Q_2}}\frac{\rt^2+1}{\medtilde{F}_0(\tilde{r})\tildespace}d\tau^2
+ \sqrt{Q_1 Q_2} \; \medtilde{\Lambda} \left( \frac{d\tilde{r}^2}{\tilde{r}^2+1}+ d\theta^2 \right)
+\frac{\sqrt{Q_1 Q_2}}{\medtilde{\Lambda} \tildespace}\sin^2\theta \left( d\varphi_1  - \frac{2 \, d\tau}{b^2 R_y}   \right)^2 \cr
&&{} 
+\frac{\sqrt{Q_1 Q_2}}{\medtilde{\Lambda} \tildespace}\Ft_1(\rt)\cos^2\theta \left( d\varphi_2 - \frac{\Ft_0(\rt)}{\Ft_1(\rt)\tildespace} \frac{dv}{R_y} + \frac{1}{\Ft_1(\rt)\tildespace} \frac{2 \, d\tau}{b^2 R_y} \right)^2 \label{eq:6dmetric2} \\
&&{} 
+ \sqrt{Q_1 Q_2} \; \medtilde{\Lambda} \frac{\Ft_0(\rt)}{\Ft_1(\rt)\tildespace} \rt^2 \left( \frac{dv}{R_y} -\frac{1}{\Ft_0(\rt)\tildespace} \frac{2 \, d\tau}{b^2 R_y}\right)^2 . \nonumber
\eea
At $\rt=0$ we have $\Ft_{0}(\rt) \to 1$, $\Ft_{1}(\rt) \to 1$ and $\medtilde{\Lambda}\to 1$, and the term on the final line combines with the $d\rt^2$ term to describe the smooth shrinking of an S$^1$ at the center of a local $\IR^2$.

The \adstwo\ limit of the matter fields can be similarly derived; since this is a straightforward implementation of the above procedure, we omit the details.

\subsection{AdS$_3$ and AdS$_2$ perspectives}

From the metric \eq{eq:6dmetric1} one can read off that in the AdS$_2$ limit the $J_R$ angular momentum has gone to zero, while the solution remains non-trivial. This indicates that the internal structure deep inside the core of the solutions indeed fits inside the \adstwo\ throat, while the $J_R$ angular momentum does not survive this limit. Thus for different values of $a/b$ in the starting solution, we have the same representative in the \adstwo\ limit.

Let us compare and contrast the above AdS$_2$ limit with a more naive $a\to 0 $ limit. If one does not rescale coordinates as in \eq{eq:ads2-limit-6D}, but rather holds $r$, $t$ fixed and sends $a \to 0$, instead of the superstratum metric \eq{eq:6dmetric1} one obtains the extremal black hole solution with a large horizon~\cite{Bena:2016ypk}. This can be interpreted as the solution effectively seen by an observer who remains at a fixed depth of the extremal BTZ throat (measured from a fixed reference far from the black hole), while the total depth of the throat is taken longer and longer.

By contrast, the limit defined above can be interpreted as the solution effectively seen by an observer deep inside the throat, as the length of the throat is taken longer and longer. From such an observer's point of view, the original asymptotic \adsthree\ region goes to infinity as the limit is taken, such that the asymptotics of the solution become those given in Eq.\;\eq{eq:6dmetric1-asym}.

\section{Excitations of asymptotically AdS$_2$ superstrata}
\label{sec5}

In this section we show that the asymptotically AdS$_2$ solutions constructed in the previous section admit an infinite tower of finite-energy non-BPS normalizable excitations. 
The results in this section are obtained for the family of $(1,0,n)$ superstratum solutions, where the wave equation for minimally coupled massless scalar fields is separable \cite{Bena:2017upb}. However, we expect that the existence of towers of finite-energy excitations is a general feature of all asymptotically AdS$_2$ microstate geometries with a smooth IR cap. From the perspective of ten-dimensional Type IIB supergravity compactified on T$^4$, the scalar fluctuations we consider come from traceless deformations of the internal manifold.

Our analysis involves an analytic solution for large $n$, presented in Section \ref{subsec:analyticsolve}, and a numerical solution for general $n$, presented in Section \ref{subsec:resultmodes}.
In Section \ref{subsec:WEads3perspecive} we discuss these results from the AdS$_3$ perspective. 
When glued back to \adsthree, these excitations correspond to towers of CFT$_2$ excitations whose energies are evenly spaced. Interestingly, for the solutions with the longest throats, the gap between these energies is equal to the smallest possible gap of the dual CFT$_2$.

\subsection{The minimally coupled massless scalar wave equation}
\label{subsec:reviewwaveeq}

We start by considering the asymptotically \adsthree$\times$S$^3$ $(1,0,n)$ family of superstratum solutions constructed in \cite{Bena:2016ypk} and reviewed in Section \ref{subsec:singlemodeS}.  In the D1-D5-P duality frame,  the Type IIB string-frame metric is
\be
ds_{10}^2 \:=\: \sqrt{\frac{Z_1 Z_2}{\cP}} \, ds_6^2 + \sqrt{\frac{Z_1}{ Z_2}}\, \delta^{(4)}_{ij}\, dx^i dx^j, \qquad i,j=1,\ldots 4,
\ee
where the six-dimensional metric is given in \eqref{eq:6dmetric1} and \eqref{eq:6dmetric2}. This choice of family is motivated by the fact that the wave equation of a massless minimally coupled scalar is separable and the null geodesic equations are integrable \cite{Bena:2017upb}. We consider a scalar deformation of the T$^4$ metric,
\be 
\delta^{(4)}_{ij}\, dx^i dx^j \, \rightarrow \, \left(\delta^{(4)}_{ij} + h_{ij} \right) \, dx^i dx^j.
\ee
The equations of motion at first order in $h_{ij}$ require that $h_{ij}$ be a minimally coupled scalar fluctuation in six dimensions (see for example Appendix B of \cite{Bombini:2017sge}), obeying the six-dimensional Klein-Gordon equation:
\begin{equation}
\frac{1}{\sqrt{-\det g}} \: \partial_M \left( \sqrt{-\det g}\: g^{MN} \partial_N \, h_{ij} \right) \:=\: 0 \,.
\end{equation}
From a six-dimensional perspective the indices $i,j$ label different scalar fields; we will take any one of these and denote it as $\Phi$ for the rest of the section. One can either directly compute the wave equation from the asymptotically \adstwo\ superstratum metric \eqref{eq:6dmetric1} or use the wave equation for the $(1,0,n)$ family of asymptotically \adsthree\ superstrata derived in~\cite{Bena:2017upb}, and take the AdS$_2$ limit of this wave equation. Both methods are equivalent. For later convenience we exhibit here the second method, recalling the main results of \cite{Bena:2017upb} in the process.

We separate variables as\footnote{Note that this separation ansatz appears slightly different to that of \cite{Bena:2017upb}, because our six-dimensional coordinates ($\{t,v\}\equiv \{t ,t+y\}$) are different from those of \cite{Bena:2017upb} ($\{u,v\}\equiv \frac{1}{\sqrt 2}\{t-y,t+y\}$). For a discussion on these two choices of coordinates, see Appendix B of \cite{Bena:2016agb}.}
\begin{equation}
\Phi = K(r) S(\theta)e^{i\left( \frac{1}{R_y} \omega \;\! t+\frac{1}{R_y}p \;\! v +q_1 \varphi_1 + q_2 \varphi_2  \right)}.
\label{eq:modeprofile}
\end{equation}
In the background of an asymptotically \adsthree\ $(1,0,n)$ superstratum, the wave equation separates \cite{Bena:2017upb} into:
\bea
\frac{1}{r} \partial_r \left( r \left( r^2 + a^2 \right) \partial_r \, K(r)  \right) + 
\left( \frac{a^2\left( \omega + p +q_1\right)^2}{r^2+a^2}-\frac{a^2\left(  p +q_2\right)^2}{r^2}\right) K(r)  && 
\label{eq:radialeq}\\
+ \frac{b^2 \omega \left( a^2 (\omega+ 2 p) + F_0(r) (2 a^2 \,q_1 + (a^2 +\frac{b^2}{2}) \, \omega)\right)}{2a^2 (r^2+a^2)}K(r)  &=&  \lambda \, K(r)  \,,\nonumber
\eea
\vspace{-3mm}
\bea
\frac{1}{\sin \theta \cos \theta} \partial_\theta \left( \sin \theta \cos \theta \, \partial_\theta \,S(\theta) \right) - \left( \frac{q_1 ^2 }{\sin^2 \theta} + \frac{q_2 ^2 }{\cos^2 \theta} \right) S(\theta) &=& -\lambda S(\theta)\,,
\label{eq:Seqads2}
\eea
where
\begin{equation}
F_0(r) \;\equiv\; 1 - \frac{r^{2 n }}{(r^2+a^2)^{n}} \,.
\end{equation}

To describe fluctuations of the asymptotically AdS$_2$ superstrata that we have constructed in Section \ref{sec4} we take the same limit as \eqref{eq:ads2-limit-6D}, rescaling $\omega$ appropriately:
\begin{equation}
r ~=~ a \tilde{r} \,, \qquad t ~=~ \frac{\tau}{a^2} \, , \qquad \omega~=~ a^2  \tilde{\omega}\, ; \quad\qquad
a ~\to~ 0 \qquad \mathrm{with} ~~ \tilde{r} \,, \tau \,, v \,,p\, ,b\,, q_1 \,, q_2 ~~ \mathrm{fixed} \,.
\label{eq:ads2-limit-modes}
\end{equation}
The scalar wave equation for the mode
\begin{equation}
\Phi = K(\tilde{r}) S(\theta)e^{i\left(\frac{1}{R_y} \tilde{\omega} \;\! \tau+\frac{1}{R_y} p \;\! v +q_1 \varphi_1 + q_2 \varphi_2  \right)}
\label{eq:modeprofileads2}
\end{equation}
of course remains separable in our AdS$_2$ limit. The angular part of the wave equation \eqref{eq:Seqads2}  remains the same, and the radial wave equation becomes 
\begin{equation}
\begin{split}
\frac{1}{\tilde{r}} \partial_{\tilde{r}} \left( \tilde{r} \left( \tilde{r}^2 + 1 \right) \partial_{\tilde{r}} \, K(\tilde{r}) \right) &+ 
\left( \frac{\left( p+q_1\right)^2}{\tilde{r}^2+1}-\frac{\left(  p +q_2\right)^2}{\tilde{r}^2}\right) K(\tilde{r}) \\
&{}\qquad + \frac{b^2  \tilde{\omega} \left(p + \tilde{F}_0(\tilde{r})\, \left(q_1+\frac{b^2}{4}\tilde{\omega}\right) \right)}{ \tilde{r}^2+1}K(\tilde{r}) ~=~ \lambda \,K(\tilde{r})\,,
\label{eq:radialeqads2}
\end{split}
\end{equation}
where $\tilde{F}_0(\tilde{r})$ is defined in \eq{eq:F0tilde}.
The angular equation \eqref{eq:Seqads2} is solvable and there is only one branch of well-defined solutions:
\begin{equation}
S(\theta) \:\propto\:  \left(\sin\theta\right)^{|q_1|}\,  \left(\cos\theta\right)^{|q_2|}\, _2 F_1 \left(-s, 1+ s +|q_1|+|q_2|; |q_2|+1; \cos^{2} \theta \right), 
\label{eq:angularSolution1}
\end{equation}
where $s$ and $l$ are given by
\begin{equation}
\lambda \,=\, l (l +2)\,, \qquad s=\frac{1}{2}\Big(|\,l+1| -1 -|q_1|-|q_2|\Big).
\label{eq:ldef}
\end{equation}
The solution is regular at $\cos^2\theta = 1$ if and only if $s$ is a non-negative integer. Consequently, the angular wave function is regular when
\begin{equation}
~~ |\,l+1| \;\geq\; 1 + |q_1| + |q_2|\,, \qquad\quad q_1,\,q_2 ,\, l, \, s\,\in\, \mathbb{Z} \,, \quad s \geq 0 \,.
\label{eq:angularconstraint}
\end{equation}
This significantly constrains the possible values of $q_1$, $q_2$ and $l$. For instance the value $l=-1$ which corresponds to tachyonic perturbations of AdS$_2$ ($\lambda=-1$) is not allowed.

The radial wave equation \eqref{eq:radialeqads2} does not appear to be analytically solvable for general $n$. In what follows we shall therefore use a combination of numerical and analytical arguments to show that it has an infinite tower of finite-energy normalizable bound-state solutions.

\subsection{Constructing finite-energy solutions}
\label{subsec:solutionprocedure}

We perform a change of variables in order to map the infinite radial direction to a segment,
\begin{equation}
z \;\equiv\; \frac{\tilde{r}^2}{1+\tilde{r}^2} \quad \iff \quad \tilde{r} \;\equiv\; \sqrt{\frac{z}{1-z}} \;, \qquad z \in \left[0,1\right). 
\end{equation}
The radial wave equation then becomes
\begin{equation}
\begin{split}
 \partial_z \left(z \,\partial_z\,K(z)\right) \:&+\: \frac{1}{4(1-z)} \bigg[ \left( p+q_1\right)^2 - \frac{1}{z} \left(  p +q_2\right)^2\\
 &  \qquad \qquad \quad  +  b^2 \tilde{\omega} \left( p + \left(1- z^{n} \right)\, \left(q_1+\frac{b^2}{4}\tilde{\omega}\right)  \right) - \frac{l (l +2)}{ 1-z}  \bigg] \, K(z) \:=\: 0.
\label{eq:radialeqsegment}
\end{split}
\end{equation}

We first investigate the behavior of the solutions to this equation around the ends of the segment ($z=0$ and $z=1$) to check whether there are any obvious restrictions to constructing bound states:  
\begin{itemize}
\item When $p+q_2 \neq 0$, near $z =0$ a solution of \eqref{eq:radialeqsegment} must satisfy
\begin{equation}
z\, \partial_z \left(z \,\partial_z\,K(z)\right) \:-\:  \left( \frac{p +q_2}{2}  \right)^2 K(z)  \:=\: 0.
\end{equation}
The only branch of regular solutions is
\begin{equation}
K(z) \:\propto\:  z^{\frac{| p +q_2 |}{2}} \:\underset{\tilde{r}\rightarrow 0}{\sim}\: \, \tilde{r}^{| p +q_2 |} .
\label{eq:behaviorat0}
\end{equation}
Consequently, regular solutions necessarily go to 0 when $\tilde{r} \rightarrow 0$. This is expected, since the spacetime caps off smoothly at this location.

\item When  $p+q_2 =0$ one must consider the next-to-leading-order term in \eqref{eq:radialeqsegment}. The resulting equation also has a regular branch of solutions at $z = 0$. The main difference is that these regular solutions remain finite at  $z = 0$. 

\item Near $z = 1$ and for $l \neq \{0,\,-2\}$, Equation \eqref{eq:radialeqsegment} becomes:
\begin{equation}
\partial_z \left(z \,\partial_z\,K(z)\right) \:-\: \frac{l (l +2) }{4\left(1-z\right)^2}  K(z)  \:=\: 0.
\end{equation}
The branch of non-diverging solutions is
\begin{equation}
K(z) \:\propto\:(1-z)^{\frac{1+\nu}{2}} \, _2 F_1 \left( \frac{\nu}{2},  \frac{\nu}{2}; \nu; (1-z)\right) \underset{z\rightarrow 1}{\sim} (1-z)^{\frac{1+\nu}{2}}  \:\underset{\tilde{r}\rightarrow \infty}{\sim} \: \frac{1}{\tilde{r}^{1+\nu}},
\label{eq:behaviorat1}
\end{equation}
with
\begin{equation}
\nu \equiv | 1 + l \,|.
\label{eq:nu-def}
\end{equation}

These solutions decay at infinity for any value of $l$. In order to check whether they correspond to normalizable or non-normalizable modes one has to check whether the energy of this field is finite. The Hamiltonian density is composed of terms of the form $\sqrt{-g} \, g^{MN}\,  \partial_M \Phi \, \partial_N \Phi $ (no sum over the indexes). 
The most important terms when  $\tilde{r} \rightarrow +\infty $ decay as
\be
\sqrt{-g} \, g^{MN}\,  \partial_M \Phi \, \partial_N \Phi \, = \, \underset{\tilde{r}\rightarrow \infty}{\cO} \left( \frac{1}{\tilde{r}^{1+2 \nu}} \right).
\label{eq:finiteenergycond}
\ee
From Eq.\;\eqref{eq:angularconstraint} we have that $l\neq -1$ and so $\nu\ge 1$. Consequently, the scalar field bound states have finite energy.

\item When $l=\{0,\,-2\}$, the behavior of $K$ at $z=1$ is dictated by the next-to-leading-order term in $\frac{1}{1-z}$ of equation \eqref{eq:radialeqsegment}. One can show that this equation admits square-integrable finite-energy solutions when $\tilde{r} \rightarrow \infty$ that decay as $K(\tilde{r}) \underset{\tilde{r}\rightarrow \infty}{\sim} \tilde{r}^{-2}$.
\end{itemize}
These two steps do not prove the existence of bound-state solutions. However, they are necessary conditions that ensure that there are no remaining obvious obstructions to building bound-state solutions. When $p+q_2\neq 0$, if we find solutions of \eqref{eq:radialeqsegment} that go to 0 at $z = 0$ and $z = 1$, then these solutions will behave as \eqref{eq:behaviorat0} and \eqref{eq:behaviorat1} at the boundaries and will be regular finite-energy excitations. When $p+q_2=0$, we have the same features but $K$ can take a non-zero finite value at $z=0$.

\subsection{Analytic bound-state solutions for large $n$}
\label{subsec:analyticsolve}

We now analytically solve the wave equation \eqref{eq:radialeqsegment} in a $(1,0,n)$-superstratum background with $n \gg 1$, in a 1/$n$ expansion. For that purpose, we decompose the wave equation as
\begin{equation}
\cL\[\tilde{\omega}\]  \, K(z) \:-\: \frac{z^{n}}{1-z} \,\mathcal{E} \[\tilde{\omega}\]   \,K(z) \:=\: 0, 
\label{eq:WElimitnlarge}
\end{equation}
where
\begin{equation}
\begin{split}
\cL\[\tilde{\omega}\]  &\:\equiv\: \partial_z \left(z \,\partial_z\,\right) \:+\: \frac{1}{4(1-z)} \bigg[ \left( p+q_1+\frac{b^2 \tilde{\omega} }{2}\right)^2 - \frac{\left(  p +q_2\right)^2}{z} 
  - \frac{l (l+2)}{ 1-z}  \bigg], \\
\mathcal{E} \[\tilde{\omega}\]   & \: \equiv \: \frac{b^2 \tilde{\omega} \, \left(4 q_1+b^2\tilde{\omega}\right)}{16}\,,
\label{eq:WElimitdef}
\end{split}
\end{equation}
and look for regular solutions with
\begin{equation}
K(0) \:=\: \text{const. }, \qquad K(1) \:=\: 0.
\label{K-boundary}
\end{equation}

\begin{figure}[t]
\centering
\includegraphics[scale=0.45]{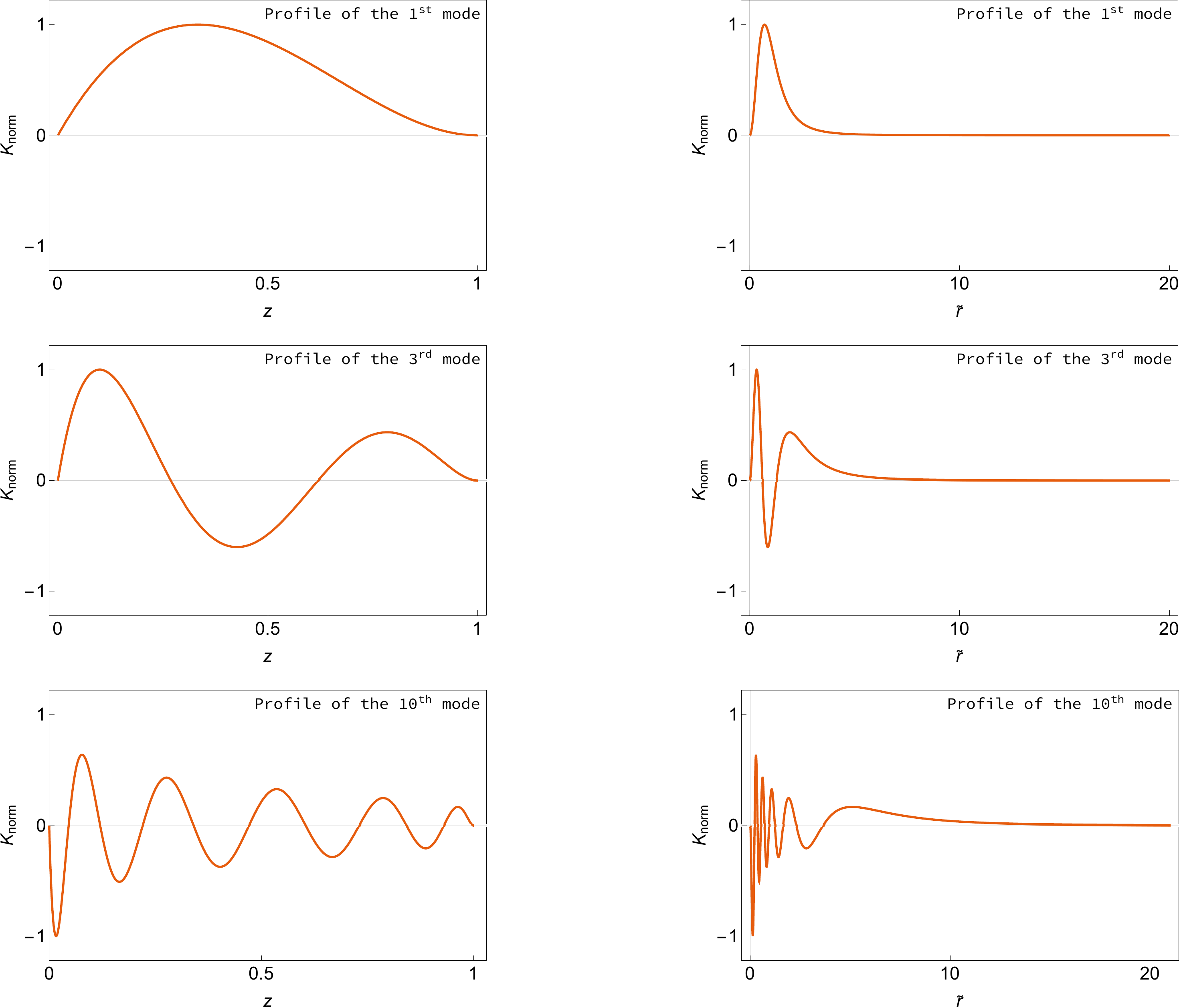}
\caption{Analytic radial wavefunctions at large $n$ for several scalar excitation modes of an asymptotically AdS$_2$ $(1,0,n)$ superstratum with $b =  1$, $n \gg 1$,  $p = q_1 = q_2 = 1$ and $l=2$. The same wavefunctions are plotted as functions of $z$ (left) and $\tilde{r}$ (right), where $z=\frac{\tilde{r}^2}{\tilde{r}^2+1}$.}
\label{fig:modesanalytic}
\end{figure}

The details of the method we use to solve this equation are given in Appendix \ref{app:WEresolution}. We show there that the only condition for having bound-state solutions is to impose $l \neq -1$ as required by \eqref{eq:angularconstraint}. For any other value of $l$, we have found a tower of excitation modes $K_N(z)$ labeled by a mode number $N \in \mathbb{N}$. The regular solutions of \eqref{eq:WElimitnlarge} are
\begin{equation}
K_N (z) \: = \: \kappa_N \:\! \left( 1 -z \right)^{\frac{1+| 1+l | }{2}}  z^{\frac{| p + q_2 | }{2}} \left[  \sum_{j=0}^N (-1)^j \, \binom{N}{j} \, \frac{\left(N +1+| 1+l | + | p + q_2 | \right)_j}{\left(1 +| p + q_2 |\right)_j} \, z^j 
+  
{\cO} \! \left(\frac{1}{n^{\nu}}\right) \right] \! ,
\label{eq:analySol1}
\end{equation}
where $(k)_j \,\equiv\, \prod_{m=0}^{j-1} (k+m)$ and $\kappa_N$ is a normalization constant. There are two possible values of $\tilde{\omega}$ for the function $K_N$ to be a solution of \eqref{eq:WElimitnlarge}. Both sets of frequencies describe the same wavefunctions, so as usual we restrict attention to the positive frequencies,
\begin{equation}
\tilde{\omega}_N \,=\, \frac{2}{b^2} \Bigl[\,2 N +1+| 1+l | + | p + q_2 | -\left(p+q_1\right) \,\Bigr] \,\:
 + \: 
{\cO} \left(\frac{1}{n^{\nu}}\right).
\label{eq:analypuls1}
\end{equation}

The leading-order term of the $\frac{1}{n}$-expansion in \eqref{eq:analySol1} captures all the features of the wavefunction. Indeed, the behaviors at $z=0$ and $z=1$ depicted in \eqref{eq:behaviorat0} and \eqref{eq:behaviorat1} are explicit in \eqref{eq:analySol1}. This proves the existence of solutions regular at both boundaries. One can re-express the modes $K_N$ in the radial variable $\tilde{r}$ using $z=\frac{\tilde{r}^2}{\tilde{r}^2+1}$. The mode profiles are depicted in Fig.~\ref{fig:modesanalytic}.

The polynomial of order $N$ in \eqref{eq:analySol1} determines the number of oscillations of the wavefunction (one can explicitly show that the polynomial has exactly $N$ roots in the range  $0< z < 1$). Much as for solutions to the Schr\"odinger equation, the lowest mode of the radial wave function has no nodes, the next one has one node, etc. One can see both from Fig.~\ref{fig:modesanalytic} and from the form of the solution that the excitations are localized in the cap and decay rapidly as one goes up the throat. We also note that, despite the complexity of the equations, the frequencies $\tilde{\omega}_N$ are linear in the mode number $N$.

Thus, we have shown that in the large $n$ limit, the asymptotically AdS$_2$ (1,0,$n$)-superstratum solutions we have constructed support an infinite tower of finite-energy non-BPS excitations. We will now investigate the same issue for arbitrary $n$ using numerical methods.

\subsection{Numerical bound-state solutions for arbitrary $n$}
\label{subsec:resultmodes}

We now describe the main steps of the procedure we use to solve Equation \eqref{eq:radialeqsegment} numerically (using {\it Mathematica}), as follows:
\begin{itemize}
\item We fix particular values for $\{ n,p,q_1,q_2, l \}$. The remaining variable is the frequency $\tilde{\omega}$.
\item When we imposed directly on $K(z)$ the Dirichlet boundary condition $K(0) = K(1) = 0$, this led the numerics to return the trivial solution $K(z) = 0$ everywhere. To evade this problem, we instead impose Dirichlet boundary conditions $K(0) = 0$ and $|K(\frac{1}{2})|=1$. Since we do not expect the solution to have a node exactly at $z = \frac{1}{2}$, this boundary condition fixes the overall normalization. 
\item We then fine-tune the value of $\tilde{\omega}$ to find the values for which $K$ goes to 0 when $z\rightarrow 1$. 
\item For each configuration $\{ n,p,q_1,q_2,l\}$ we have studied, we find a discrete set of positive $\tilde{\omega}$ for which $K$ vanishes at $z = 1$. This set of  positive frequencies $\tilde{\omega}_N$ characterizes the tower of non-supersymmetric excitations of our solutions.

\begin{figure}[t]
\centering
\includegraphics[scale=0.65]{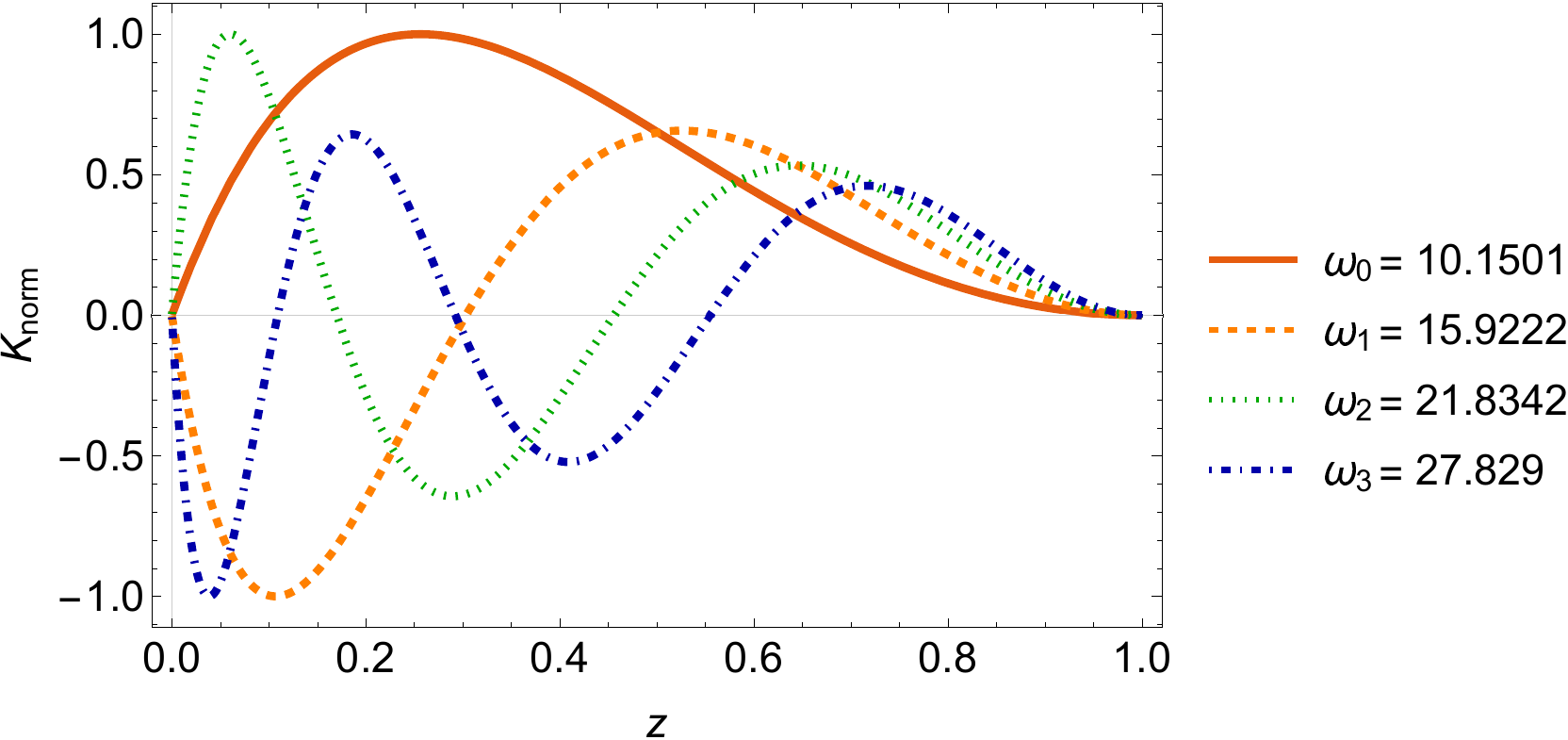}
\caption{Numerically obtained radial wavefunctions of the first four excitation modes of an asymptotically \adstwo\ (1,0,1) superstratum with $b = 1$, $p = q_1 = q_2 =1$ and $l= 2$, plotted as functions of $z$.}
\label{fig:firstmodesex}
\end{figure}

\end{itemize}
We illustrate our procedure with a particular example:\footnote{In principle, the value of $b$ should be chosen to ensure that $Q_3$ in \eq{eq:Q3-val-SS} satisfies \eq{eq:ads2-regime-SS} (recall $a$ has already taken to zero here),  however we shall simply take $b=1$ for the purpose of plotting the results. From \eq{eq:WElimitdef} the physics of the modes depends only on the combination $b^2 \tilde\omega$, so one can easily rescale as desired.}
\begin{equation}
\begin{split}
&n \, =\, 1 \,, \qquad b \,=\, 1 \,, \qquad p \, =\, q_1 \, =\, q_2 \, =\, 1 \,, \qquad l \,=\,2 \,.
\end{split}
\end{equation}
The equations governing bound states for this choice of parameters are 
\begin{equation}
\begin{split}
&\partial_z \left(z \,\partial_z\,K_N\right) \:-\: \left[ \frac{1}{z} +\frac{2}{(1-z)^2} - \frac{\tilde{\omega}_N}{4(1-z)} -\frac{\tilde{\omega}_N}{4} \left( 1+\frac{\tilde{\omega}_N}{4}\right)\right] K_N \:=\: 0, \\
&\text{with  } K_N(0) \:=\: K_N(1) \:=\: 0.
\label{eq:radialeqsegmentex}
\end{split}
\end{equation}
We apply the procedure detailed above. We find a discrete set of values of $\tilde{\omega}$ for which $K(z)$ is regular at the boundaries. Figure\;\ref{fig:firstmodesex} shows the radial wave functions for the first four modes of the tower in the $z$-coordinate system. 

\begin{figure}[t]
\centering
\includegraphics[scale=0.55]{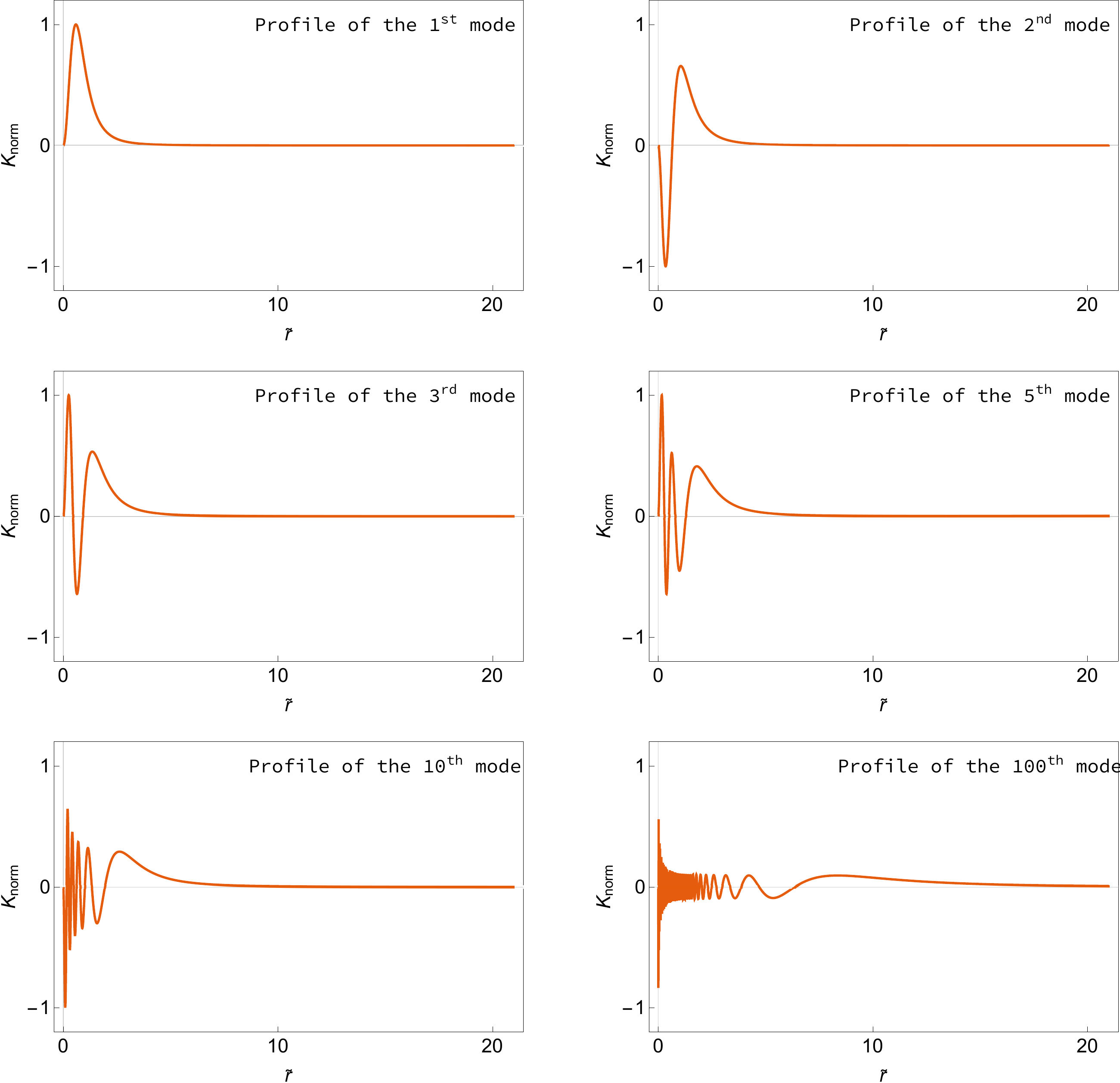}
\caption{Numerically obtained radial wavefunctions of several different excitation modes of an asymptotically \adstwo\ (1,0,1) superstratum solution with $b =1$, $  p = q_1 = q_2 =1$ and $l= 2$, plotted as functions of $\tilde{r}$.}
\label{fig:modesradialex}
\vspace{3mm}
\end{figure}

The plots in Fig.\;\ref{fig:modesradialex} show the radial component of the modes in the radial coordinate $\tilde{r}$. Their features are very similar to the ones found analytically at large $n$, shown above in Fig.\;\ref{fig:modesanalytic}. In particular, the energy grows approximately linearly with the mode number and the excitations are localized near the IR cap and decay very quickly at large $\tilde{r}$, even for high-energy excitations. 

\begin{figure}
\centering
\includegraphics[scale=0.75]{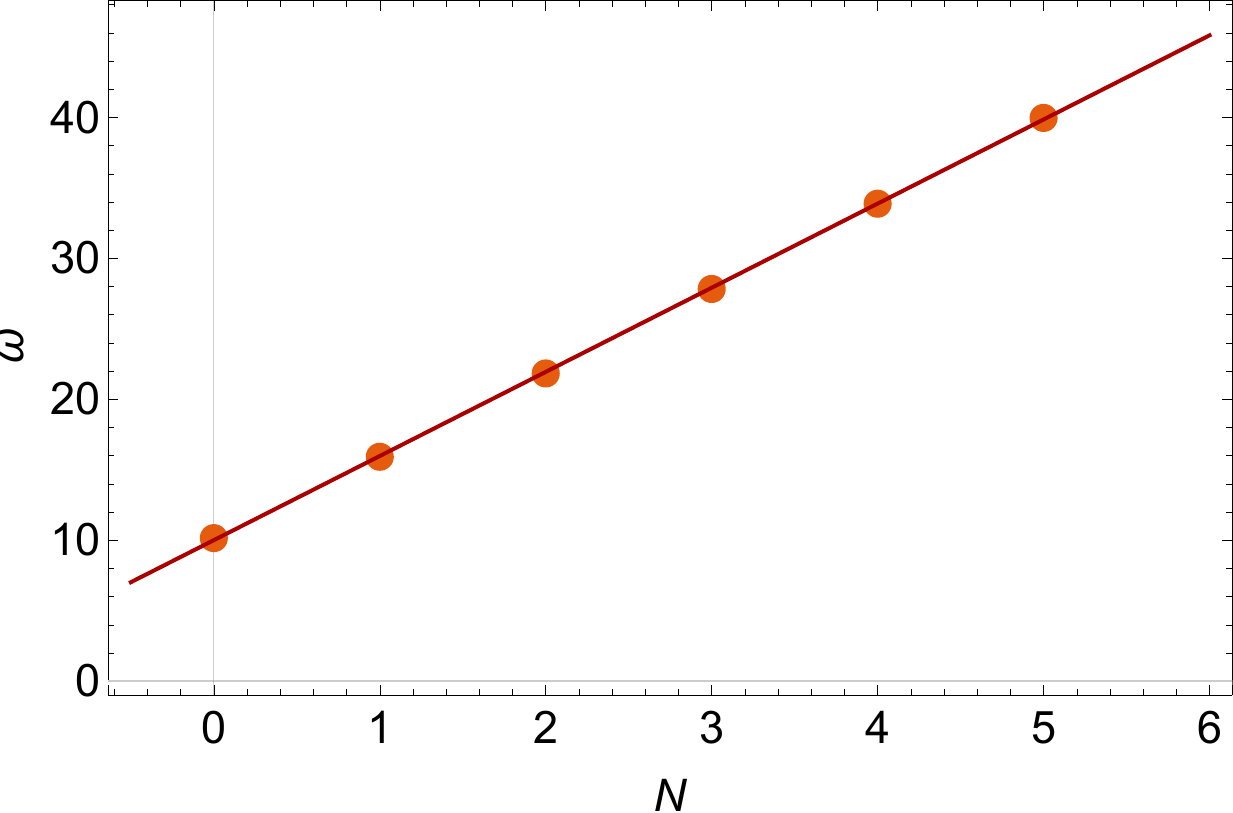}
\caption{The frequencies of modes with number $N$, and the linear fitting function given in \eqref{eq:omegalaw}.}
\label{fig:omegalaw}
\end{figure}

Interestingly, the frequencies have approximately the same linear dependence on the mode number, $N$, as that found at large $n$ in Eq.\;\eqref{eq:analypuls1} (see Fig.\;\ref{fig:omegalaw}):
\begin{equation}
\tilde{\omega}_N \:\simeq\: 5.97 \,(\, N \,+\, 1.67\,), \qquad N \in \mathbb{N}.
\label{eq:omegalaw}
\end{equation}
We repeated the numerical procedure for different values of $\{ n,p,q_1,q_2,l\}$, and we obtained similar results to those shown here.

We thus see that the existence of an infinite tower of excitations can be established analytically at  large $n$ and numerically for finite $n$. It is remarkable how similar the mode profiles are for large $n$ and for $n=1$ (compare Fig.~\ref{fig:modesanalytic} with Figs.~\ref{fig:firstmodesex} and \ref{fig:modesradialex}). It is also interesting that the frequencies depend linearly on the mode number, both at large $n$ \eqref{eq:analypuls1} and at finite $n$ \eqref{eq:omegalaw}, despite the rather complicated form of the equations we are solving.

\subsection{An AdS$_3$ perspective}
\label{subsec:WEads3perspecive}

In previous subsections we studied scalar excitations of asymptotically AdS$_2$ superstrata. 
It is also interesting to solve the wave equation of the corresponding asymptotically AdS$_3$ superstrata and to examine the properties of the modes from the perspective of an AdS$_3$ observer. For that purpose, we consider the family of asymptotically AdS$_3$ $(1,0,n)$ superstrata and the wave equation of a scalar field in this background \eqref{eq:modeprofile}--\eqref{eq:Seqads2}.

\subsubsection{Excitation modes of asymptotically AdS$_3$ solutions}

We recall that the mode profile is 
\begin{equation}
\Phi = K(r) S(\theta)e^{i\left(\frac{1}{R_y} \omega \,t+\frac{1}{R_y} p \,v +q_1 \varphi_1 + q_2 \varphi_2  \right)},
\end{equation}
where $K(r)$ and $S(\theta)$ satisfy the radial and angular equations \eqref{eq:radialeq} and \eqref{eq:Seqads2}. The solutions of the angular equation are still given by \eqref{eq:angularSolution1} and \eqref{eq:ldef}.
First, we perform a similar change of variables in order to map the infinite radial direction to a segment,
\begin{equation}
\hat{z} = \frac{r^2}{a^2+r^2} \iff r =a\, \sqrt{\frac{\hat{z}}{1-\hat{z}}}, \qquad \hat{z} \in \left[0,1\right).
\end{equation}
The radial wave equation \eqref{eq:radialeq} becomes
\begin{equation}
\hat{\cL}\[\omega\]  \, K(\hat{z}) \:-\: \frac{\hat{z}^{n}}{1-\hat{z}} \,\hat{\mathcal{E}} \[\omega\]   \,K(\hat{z}) \:=\: 0, \\
\label{eq:WElimitnlarge2}
\end{equation}
with the boundary condition $K(0) \:=\: K(1) \:=\: 0,$ and where we have defined
\begin{equation}
\begin{split}
\hat{\cL}\[\omega\]  &\:\equiv\: \partial_{\hat{z}} \left(\hat{z} \,\partial_{\hat{z}}\,\right) \:+\: \frac{1}{4(1-\hat{z})} \bigg[ \left( p+q_1+(1+\hat{b}^2)\,\omega \right)^2 - \frac{1}{\hat{z}} \left(  p +q_2\right)^2 - \frac{l (l +2)}{ 1-\hat{z}}   \bigg], \\
\hat{\mathcal{E}} \[\omega\]   & \: \equiv \: \frac{\hat{b}^2 \omega \, \left(2q_1+(1+\hat{b}^2)\,\omega\right) }{4},\\
\hat{b}&\:\equiv\: \frac{b}{\sqrt{2} \,a}.
\end{split}
\end{equation}
This equation looks similar to the radial wave equation in the asymptotically AdS$_2$ background \eqref{eq:WElimitnlarge} and \eqref{eq:WElimitdef}. 

As in Section \ref{subsec:analyticsolve}, we take $n$ to be large and work to leading order in $1/n$.
The only condition for having finite-energy excitations is to impose $l \neq -1$. 
For any other value of $l$, we have found a tower of excitation modes $K_N(\hat{z})$ labeled by a mode number $N \in \mathbb{N}$:
\begin{equation}
\begin{split}
K_N (\hat{z}) \: = \: \kappa_N \, \left( 1 -\hat{z} \right)^{\frac{1+| 1+l | }{2}}  \, \hat{z}^{\frac{| p + q_2 | }{2}} \,  & \left[ \: \sum_{j=0}^N (-1)^j \, \binom{N}{j} \, \frac{\left(N +1+| 1+l | + | p + q_2 | \right)_j}{\left(1 +| p + q_2 |\right)_j} \, \hat{z}^j \right. \\
 &\left. 
\hspace{5.6cm}\, + \: {\cO} \left(\frac{1}{n^{\nu}}\right)\,\right] ,
\end{split}
\label{eq:analySolads3}
\end{equation}
where $(k)_j \,\equiv\, \prod_{m=0}^{j-1} (k+m)$ and $\kappa_N$ is a normalization constant. The wave function $K_N$ is again a solution of \eqref{eq:WElimitnlarge2} for two values of $\omega$. The tower of positive values of $\omega$ is given by
\begin{equation}
\omega_N \,=\, \frac{1}{1+\hat{b}^2} \Bigl[\,2 N +1+| 1+l | + | p + q_2 | -\left(p+q_1\right) \,\Bigr] \,\: + \: 
{\cO} \left(\frac{1}{n^{\nu}}\right).
\label{eq:analypulsads3}
\end{equation}
Furthermore, we have a relation between $1+\hat{b}^2$ and the quantized charges of the solution \cite{Bena:2017xbt}
\be
a^2 + \frac{b^2}{2} ~=~ \frac{Q_1 Q_2}{R_y^2} \qquad \Rightarrow \qquad \frac{1}{1+\hat{b}^2} ~=~ \frac{a^2 R_y^2} {Q_1 Q_2} ~= ~\frac{2\, J_R}{N_1 N_5} \,,
\ee
where $N_1$, $ N_5$ are the integer numbers of D1 and D5 branes, and where $J_R$ is dimensionless and quantized in units of 1/2 (so $J_R=1/2$ corresponds to the solution with the longest throat from the \adsthree\ perspective). One can now compute the mass gap, $\Delta$, of our perturbations, which is equal to the smallest excitation energy above the ground state, and also equal to the difference between two successive energies in the tower \eqref{eq:analypulsads3}:
\be
\Delta \,\equiv\, \frac{\omega_N - \omega_{N-1} }{R_y} \,=\, \frac{4 \, J_R}{N_1 N_5 R_y}\,.
\ee
The mass gap was previously estimated by an order-of magnitude calculation in \cite{Bena:2016agb} and by an infrared analysis of the wave equation in \cite{Tyukov:2017uig}, where it was also pointed out this gap is of the same order of magnitude as the smallest mass gap in the D1-D5 CFT. 

In fact, for non-BPS excitations of the D1-D5 CFT, the lowest gap in the theory is obtained by adding one unit of left-moving and one unit of right-moving energy to a ground state in the longest possible winding sector, of winding $N_1 N_5$, and is equal to  $\frac{2}{N_1 N_5 R_y}$. Analytically solving the wave equation allows us to also pin down the exact coefficient of the bulk mass gap and to find that the solution with the longest throat has a gap exactly equal to the CFT$_2$ gap in this `long string' sector: $\frac{2}{N_1 N_5 R_y}$.

By contrast, at the orbifold point the proposed dual CFT$_2$ states to these superstrata have strands of length one, so the mass gap at the free orbifold point in the moduli space is $\frac{2}{R_y}$. Of course, the gap is not a protected quantity and so a mismatch is both expected and in line with previous findings~\cite{Lunin:2002iz,Bena:2016agb}. However, it is remarkable (and worthy of future investigation) that in this example the gap appears to be renormalized by precisely the maximal amount $N_1 N_5$.

\subsubsection{Infinite-throat limit of the excitation modes} 

We next describe how the perturbations behave as $a\rightarrow 0$, when the length of the AdS$_2$ region increases indefinitely.
First, we rewrite the mode profiles and their corresponding frequencies to make the $a$-dependence explicit,
\begin{equation}
\Phi_N = K_N(r)\, S(\theta)\,e^{i\left(\frac{1}{R_y}\omega_N \,t+\frac{1}{R_y} p \,v +q_1 \varphi_1 + q_2 \varphi_2  \right)},
\label{eq:modeprofiletower}
\end{equation}
where
\begin{equation}
\begin{split}
K_N (r) \: = \: & \kappa_N \, \left( \frac{a^2}{r^2+a^2} \right)^{\frac{1+| 1+l | }{2}}  \,  \left( \frac{r^2}{r^2+a^2} \right)^{\frac{| p + q_2 | }{2}} \,  \times \\
& \left[ \: \sum_{j=0}^N (-1)^j \, \binom{N}{j} \, \frac{\left(N +1+| 1+l | + | p + q_2 | \right)_j}{\left(1 +| p + q_2 |\right)_j} \,  \left( \frac{r^2}{r^2+a^2} \right)^j \right.
\left. \, + \: 
{\cO} \left(\frac{1}{n^{\nu}}\right)\,\right] ,
\end{split}
\end{equation}
\begin{equation}
\omega_N \,= \,  \frac{a^2 R_y^2}{Q_1 Q_2} \left[\,2 N +1+| 1+l | + | p + q_2 | -\left(p+q_1\right) \,\right] \,\: + \: 
{\cO} \left(\frac{1}{n^{\nu}}\right),\qquad \quad
\end{equation}
and where $S(\theta)$ is given as before by \eqref{eq:angularSolution1} and \eqref{eq:ldef}.\\
\noindent $\bullet$ For an observer at the top of the \adstwo\ throat, near the AdS$_2$--AdS$_3$ gluing region, the six-dimensional coordinate system is the original one $\{ r,t,v,\theta,\phi_1,\phi_2 \}$. The limit $a \rightarrow 0$ is trivial and gives $\omega_N \rightarrow 0$. This means that the perturbations seen by such an observer are red-shifted to zero-energy perturbations. This confirms the point of view that the AdS$_3$ perspective is inappropriate to study asymptotically AdS$_2$ geometries.   \\
\noindent $\bullet$ For an observer at the bottom of the throat, we use the rescaled coordinates \eqref{eq:ads2-limit-6D}:  $\{ \tilde{r}, \tau ,v,\theta,\phi_1,\phi_2 \}$. In these coordinates, the leading terms in $1/n$ of the radial parts of the excitation modes are independent of $a$ and the frequencies depend on $a$ only through the combination $a^2 + b^2/2 \,=\, Q_1 Q_2/R_y^2$,
\begin{equation}
\tilde{\Phi}_N = \tilde{K}_N(r) S(\theta)e^{i\left(\frac{1}{R_y}  \, \tilde{\omega}_N \,\tau+\frac{1}{R_y} p \,v +q_1 \varphi_1 + q_2 \varphi_2  \right)},
\end{equation}
where
\begin{equation}
\begin{split}
\tilde{K}_N (\tilde{r}) \; =  & ~\, \kappa_N \:\! \left( \frac{1}{\tilde{r}^2+1} \right)^{\frac{1+| 1+l | }{2}}  \,  \left( \frac{\tilde{r}^2}{\tilde{r}^2+1} \right)^{\frac{| p + q_2 | }{2}} \,  \times \\
& ~~ \left[ \: \sum_{j=0}^N (-1)^j \, \binom{N}{j} \, \frac{\left(N +1+| 1+l | + | p + q_2 | \right)_j}{\left(1 +| p + q_2 |\right)_j} \,  \left( \frac{\tilde{r}^2}{\tilde{r}^2+1}\right)^j \right.
\left. \, + \: 
{\cO} \left(\frac{1}{n^{\nu}}\right)\,\right] ,
\end{split}
\end{equation}
\begin{equation}
\tilde{\omega}_N \;=\; \frac{1}{a^2+\frac{b^2}{2}}  \Big[\,2 N +1+| 1+l | + | p + q_2 | -\left(p+q_1\right) \;\! \Big] \: + \: 
{\cO} \left(\frac{1}{n^{\nu}}\right).\qquad \quad
\end{equation}
In the limit $a \to 0$, these expressions correspond precisely to the modes on top of the asymptotically AdS$_2$ $(1,0,n)$ superstrata \eqref{eq:analySol1}, \eqref{eq:analypuls1}. In other words, the perturbations of the asymptotically AdS$_3$ solutions live in the cap at the bottom of the intermediate \adstwo\ throat, and when the throat gets longer and longer they become the perturbations of an asymptotically AdS$_2$ solution as seen by an observer at the bottom of the infinite throat.

To conclude this section, the solution of the wave equation on the asymptotically AdS$_3$ solutions raises three interesting points. First, the mass gap of scalar excitations in the bulk is $\frac{4 \, J_R}{N_1 N_5 R_y}$. For the solution with the longest throat, this matches exactly the lowest mass gap of non-BPS excitations of D1-D5 CFT$_2$, and is $N_1 N_5$ times larger than the value computed at the free orbifold point of the moduli space. Second, from the perspective of an observer at the bottom of the \adstwo\ throat, the tower of excitations is the same as the tower of excitations on top of a asymptotically AdS$_2$ solution plus small corrections. Thus, there is a one-to-one mapping between the excitation modes of asymptotically AdS$_3$ and the excitation modes of asymptotically AdS$_2$ microstate geometries.

Third, the frequencies of the modes and hence the spacing between different energy levels constructed on top of these solutions depend linearly on the mode number, which is quite remarkable given the intricate form of the equations we solved to obtain these energies. A linear spectrum agrees with what expects from a CFT$_2$ on a cylinder, and it would be interesting to understand whether this is a feature of more general superstratum solutions. 

We note in passing that it has recently been argued that there is a tension between the fact that the spacing of excitations above the BPS D1-D5-P asymptotically \adsthree\ superstrata with the longest throats is of order ${1 \over N_1 N_5}$, and the fact that energy differences of non-BPS states away from the BPS bound are generically of order $e^{-S}$, where $S$ is the entropy\footnote{The entropy of non-BPS D1-D5-P black holes is given by $S = 2\pi \sqrt{N_1 N_5 N_P-J_L^2}+  2\pi \sqrt{N_1 N_5 N_{\bar P}-J_R^2}$, where $N_P = L_0 - c/24$ and $N_{\bar{P}} = \bar{L}_0-c/24$ are the left- and right-moving excitation numbers.} of the black hole~\cite{Raju:2018xue}.
However, these two facts are not in any sharp tension: above the many BPS states, different 
towers of excitations with spacings of order ${1 \over N_1 N_5}$ that are not exactly the same will generically give rise to differences in energies of non-BPS states of order $e^{-S}$ because of the large degeneracy of states.
Note that any given pair of non-BPS states with an energy difference of order $e^{-S}$ could easily lie in different topological sectors and/or different regions of parameter space, so one should not expect such an energy difference to be visible in the perturbations of a single microstate geometry.

\section{Discussion}
\label{discussion}

In this paper we have constructed two large classes of asymptotically \adstwo\ supergravity solutions, and formulated a general procedure to construct even larger classes of such solutions. We have also solved for a tower of non-supersymmetric excitations above a family of capped asymptotically \adstwo\ solutions. Returning to \adstwo\ holography, we now discuss in detail the implications of our results, especially the non-supersymmetric bound state excitations we have found, focusing on the key question: {\it What is their backreaction of these modes, and what does this imply for the dual CFT$_{1}$?} 
\vspace{8mm}

\subsection{Backreaction}
\label{subsec:backreaction}

One possibility is that the backreaction of our modes necessarily modifies the \adstwo\ UV asymptotics, and gives rise to ``running dilation'' or Nearly-\adstwo\ solutions where the volume of the compact directions grows in the UV and the  \adstwo\ throat is glued to an \adsthree, AdS$_4$, or flat spacetime.
 In global  \adstwo\ and in simple theories (such as Jackiw-Teitelboim gravity) it  has been  argued that all finite-energy perturbations induce a running dilaton that modifies the \adstwo\ UV asymptotics \cite{Maldacena:1998uz,Almheiri:2014cka}. 
However, our solutions cap off smoothly in the IR, so these arguments do not directly apply.

One might nevertheless imagine that the finite energy of any backreacted solution might mean that the UV is necessarily modified. 
However, it is known that there exist time-dependent non-extremal black hole solutions that have non-degenerate horizons in the IR and remain asymptotically \adstwo\ in the UV \cite{Castro:2008ms}\footnote{The arguments of \cite{Maldacena:1998uz,Almheiri:2014cka} do not apply to these solutions either, because of the presence of a horizon.}. Thus the mere presence of a non-trivial energy-momentum tensor in the bulk is not enough to destroy the \adstwo\ UV asymptotics. 
Indeed, if the backreacted solutions preserve the \adstwo\ UV, they would be natural candidate microstates of these non-extremal black holes.\footnote{These black holes can also be obtained from non-extremal asymptotically flat black holes by a similar limit to that defined in Section \ref{sec:general-limit}.}

Furthermore, if the backreaction of the finite energy bound-state modes necessarily modifies the \adstwo\ UV asymptotics, it gives rise to a puzzle: In such a scenario, the resulting finite length of the \adstwo\ throat would depend on the energy above extremality, and the natural expectation is that this dependence should be inverse-linear. That is, from an \adsthree\ perspective, the lowest state would still have energy ${2\over N_1 N_5}$, coming from the lowest-energy mode in the longest throat. The next state would have twice more energy from the perspective of an observer at the bottom of the throat, but the throat itself would be be shorter by a factor of two because of the inverse-linear dependence on the energy added. Hence, from the \adsthree\ perspective  the energy of the second mode would be ${4 \!\times\! {2\over N_1 N_5}}$, and so on, leading to a quadratic spectrum of excitations that does not resemble the spectrum of any CFT$_2$ known to us. This conclusion could perhaps be avoided if the dependence of the throat length on the added energy is sufficiently weak, although we have not managed to construct a credible model for this possibility.

This puzzle, together with the existence of non-extremal black holes with \adstwo\ UV asymptotics, is in our opinion a strong indication that the backreaction does not necessarily modify the \adstwo\ UV. We now scrutinize this alternative possibility and show that it passes some basic tests.

First, one should identify the physics that controls whether or not the UV is modified into a throat of finite length. We propose that generically the throat length should be controlled by a combination of the overall angular momentum $J_R$ and the contributions to $J_R$ from the topological bubbles of the solution.  The throat lengths of the asymptotically \adsthree\ superstrata studied in the previous section are indeed controlled by $J_R$, and we have seen that their energy gaps, as well as the difference between higher energy levels are $ 4 J_R \over N_1 N_5 $. As described in the previous section, this is consistent with D1-D5 CFT physics.

Before taking the strict \adstwo\ limit of the deep-scaling solutions studied in this paper, the long \adstwo\ throats and their gluing regions
correspond, from an \adstwo\ perspective, to running-dilaton solutions (where the dilaton encodes the size of a compact space transverse to \adstwo). Our method to construct asymptotically \adstwo\ solutions has been to scale into a region of parameter space where the dilaton starts running further and further out in the \adstwo\ UV, and to take the \adstwo\ limit which restricts to a locus in parameter space where the dilaton becomes asymptotically constant.

In this language, one might imagine that introducing a non-supersymmetric perturbation and keeping all the parameters fixed on this locus in parameter space, may also induce a running dilaton. However it appears by parameter counting that there should be enough freedom to re-adjust parameters to compensate the non-supersymmetric contribution to the dilaton equation of motion and re-set the dilaton to asymptote to a constant value. Hence, both in the black-hole solutions of  \cite{Castro:2008ms} and in our solutions, we expect that the parameters that control the running of the dilation in the UV are independent of the presence of a finite-energy configuration in the IR, and that finite-energy perturbations are compatible with  \adstwo\ UV asymptotics.

For the above reasons, the working hypothesis we consider for the remainder of our \adstwo-CFT$_1$ discussion is that our finite-energy excitations backreact into solutions that preserve the (constant dilaton) \adstwo\ asymptotics, and that are dual to time-dependent configurations of the CFT$_1$.

\subsection{Holographic description of the solutions and excitations}
\label{sec:disc-holo}

A fascinating question, whose answer is beyond the scope of our paper, is to investigate whether or not the finite-energy bound state excitations correspond to finite-energy excitations in the dual CFT$_1$. In~\cite{Cvetic:2016eiv} it was argued that for certain simple two-dimensional Maxwell-dilaton theories, the holographic stress tensor is identically zero whenever the UV asymptotics is \adstwo, and hence all the constant-dilaton asymptotically \adstwo\ solutions, including the time-dependent black holes of~\cite{Castro:2008ms}, have zero energy in the CFT$_1$. Since the CFT$_1$ configurations corresponding to these black holes have time-dependent VEVs, and since one usually associates time dependence with finite energy, this would appear to be a distinctly unconventional property of the CFT$_1$. However, a priori it does not appear to be ruled out.

Having said this, the asymptotically \adstwo\ superstrata descend from string theory solutions with nontrivial harmonics along the compact directions and both R-R and NS-NS potentials turned on. Thus their field content from an \adstwo\ perspective consists of a large (and possibly infinite) number of Kaluza-Klein modes, as well as several scalars and vectors, and hence is much more general than the field content of the solutions considered in~\cite{Cvetic:2016eiv}. Furthermore, our supersymmetric solutions do not have an everywhere-constant dilaton as in~\cite{Cvetic:2016eiv}, but a dilaton that asymptotes to a constant. Thus, computing the holographic stress tensor for our theories may give a different result.
A priori it seems more likely that the finite-energy bound state excitations we construct are dual to finite-energy time-dependent states of the CFT$_1$. This would fit much better with the expectation that time dependence corresponds to finite energy, and would indicate that the CFT$_1$ is a more conventional theory. 

Another reason to expect that our finite-bulk-energy bound states should be dual to finite-energy CFT$_1$ states is that we have an infinite tower of  finite-energy perturbations on top of a family of bulk solutions, and we expect to have such a tower of perturbations on top of every such supersymmetric solution. 
Hence, if the CFT$_1$ energy of all these modes were zero, there would be a dramatic overcounting problem of the CFT ground states. Of course one could argue that only $e^S$ of these states are supersymmetric and contribute to the index, while the other states do not, but this would again be a distinctly unusual situation.

Furthermore, the bulk theory also has an infinite number of black hole solutions, parameterized by an arbitrary function of one variable~\cite{Castro:2008ms} and, if these black holes correspond to ensembles of CFT$_1$ zero-energy states, then again there would appear to be a serious overcounting problem of the CFT ground states.

If the finite-energy bulk perturbations had finite CFT$_1$ energies, this would solve all these problems: the CFT$_1$ would not have an infinite entropy at zero energy and no states at finite energy, but rather a finite entropy at each energy level. Furthermore, the entropy at each energy level might be captured by the non-extremal black holes of~\cite{Castro:2008ms}. 

When the holographic CFT$_1$ has a parent holographic CFT$_2$ dual to \adsthree, our construction can used to better understand the relation between the CFT$_1$ and the parent CFT$_2$. The asymptotically \adstwo\ solutions constructed in Section~\ref{sec4} are obtained as a limit of families of asymptotically \adsthree\ solutions with increasingly longer throats and progressively decreasing $J_R$. 
The asymptotically \adstwo\ solutions are formally the $J_R=0$ members of these families.
It is therefore tempting to postulate that for every microstate of the CFT$_1$ there is a corresponding family of CFT$_2$ microstates, parameterized by the value of $J_R$. It should be understood that this applies to states with relatively low values of $J_R$, that are dual to solutions with long throats.

From the perspective of the explicit families of D1-D5 orbifold CFT$_2$ states dual to asymptotically \adsthree\ superstrata (see~\cite{Bena:2016ypk,Bena:2017xbt}), the expectation value of $J_R$ is given by half the average number of $(+,+)$ Ramond-Ramond ground state strands, so decreasing or increasing $J_R$ corresponds to changing this average. Provided that the number of $(+,+)$ strands remains small, this produces a small relative change in the physics of the other strands of the CFT$_2$.
Furthermore, it is tempting to imagine that, away from the orbifold point, the remaining strands form some kind of effective `long string' that explains the small gap in the bulk.

One could then think of this one-to-many correspondence between \adstwo\ and \adsthree\ microstates, and our \adstwo\ limit, as zooming on the information carried by the long string, which is the information encoding the bulk of the black hole entropy, and ignoring the information from the very small number of $(+,+)$ strands. These strands could then be thought of as principally encoding the information pertaining to how the \adstwo\ throat geometry is embedded inside AdS$_3$.

\subsection{Other future directions}
\label{sec:disc-future-direc}

Our results open up several other interesting directions for future work.
By computing two-point functions in our geometries, one can compute correlators in the dual CFT$_1$, similar to the four-point functions of two heavy and two light operators that have been calculated in the D1-D5 CFT \cite{Giusto:2015dfa,*Galliani:2016cai,*Galliani:2017jlg,Bombini:2017sge}.
One could then compare these quantities to related quantities computed in theories such as the SYK model and its generalizations.
%

The existence of large classes of bulk solutions that are dual to supersymmetric ground states of the CFT$_1$ implies that this theory has a large number of dimension-zero operators that transform these states into each other. These operators form a large symmetry group, as discussed in \cite{Sen:2011cn}.  This symmetry group, which is infinite in the classical limit, might manifest itself in the bulk similarly to the BMS symmetry group for flat space \cite{Bondi:1962px,*Sachs:1962wk}. It would be interesting to construct the corresponding conserved charges and to find whether they distinguish different microstate geometries corresponding to different CFT$_1$ states. It would also be interesting to investigate whether there may be any link along these lines with the `soft hair' ideas explored in~\cite{Hawking:2016msc}.

From the perspective of quantum gravity in \adstwo, it would be interesting to know the minimal field content needed to construct solutions that have the key features of the solutions presented in this work: being singular in two dimensions but smooth in higher dimensions, having an asymptotically constant dilaton, and having non-trivial features deep in the IR. Jackiw-Teitelboim gravity and its hitherto-studied extensions do not appear to have the necessary field content; finding this minimal theory would then enable an investigation of the mechanism that controls backreaction in the UV.\footnote{For an embedding of Jackiw-Teitelboim gravity in String Theory, see~\cite{Li:2018omr}.} It would be interesting to perform precision holographic investigations of our asymptotically \adstwo\ solutions using Kaluza-Klein holography~\cite{Skenderis:2006uy}, and also to investigate connections with the recent developments in describing string worldsheet physics of black hole microstates~\cite{Martinec:2017ztd,Martinec:2018nco}.

To settle the important question of backreaction it is desirable to construct fully backreacted asymptotically \adstwo\ non-extremal smooth microstate geometries. There has been significant recent progress in constructing non-extremal microstate solutions~\cite{Bossard:2014ola,Bena:2015drs,Bena:2016dbw,Bossard:2017vii}, and this technology appears to be the most promising avenue for this future line of investigation.

We believe that the solutions we have constructed and the physics they reveal offer a useful perspective on the rich and continuously evolving subject of quantum gravity and holography in \adstwo, and we look forward to developing this subject further in the future.

\vspace{8mm}

\noindent {\bf Acknowledgements:} We thank Jan de Boer, Billy Cottrel, \'Oscar Dias, Monica Guic\u a, David Kutasov, Finn Larsen, Oleg Lunin, Emil Martinec, Miguel Paulos, Ioannis Papadimitriou, Andrea Puhm, Rodolfo Russo, Kostas Skenderis, Marika Taylor and Nick Warner for useful discussions. This work was supported in part by the ANR grant Black-dS-String ANR-16-CE31-0004-01. The work of PH is supported by an ENS Lyon grant. The work of DT is supported by a Royal Society Tata University Research Fellowship.

\vspace{3mm}

\begin{appendix}

\section{BPS equations in five and six dimensions}
\label{app:BPS-equations}

\subsection{Five-dimensional U(1)$^4$ supergravity}
\label{app:5dsugra}

We will review the $\cN = 1$, five-dimensional supergravity coupled to four $U(1)$ gauge fields \cite{Gutowski:2004yv,*Gauntlett:2004qy,*Bena:2004de,Giusto:2012gt,*Vasilakis:2012zg,*Bena:2013ora}. This theory can obtained from an eleven-dimensional supergravity reduced on a $T^6$. The non-zero structure constants $C_{IJK}$ which occur after truncation down to five dimensions are
\be
C_{IJK} ~=~{} \lvert \epsilon_{IJK} \rvert \, ,\quad~~  I,J,K=1,2,3 ~; \qquad~~  C_{344} = -2\,.
\ee
The timelike-supersymmetric field configurations have a conformastationary metric in eleven dimensions given by
\begin{equation}
\begin{split}
ds_{11}^2 ~=~{}  -&\frac{1}{(Z_3 \cP)^{2/3}} \,(dt +  \kk)^2 \,+\,  (Z_3 \cP)^{1/3}\, ds_4^2 \\
\,+& \sum_{i=1}^3  \frac{C_{iJK} Z_J Z_K}{2\, (Z_3 \cP)^{2/3}} \, dw_i d\bar{w}_i   \,+\, \frac{C_{4JK} Z_J Z_K}{2\, (Z_3 \cP)^{2/3}} \,\left( dw_1 d\bar{w}_2 \,+\, dw_2 d\bar{w}_1\right),
\end{split}
\end{equation}
where $\cP \equiv \frac{1}{2} C_{3JK} Z_J Z_K =  Z_1 Z_2 - Z_4^2$, $Z_I$ and $\kk$ are respectively four functions and a one-form taking values in this four-dimensional space, and $w_i$ are the three complex coordinates of the three 2-tori composing the $T^6$. The four-dimensional base space associated to the metric $ds_4^2$ must be hyperk\"{a}hler to preserve supersymmetry. Supersymmetry also requires the three-form potential to be:
\begin{equation}
\cA ~=~{}  \sum_{j=1}^3 \left( A^{(j)} \wedge \frac{dw_j \wedge d\bar{w}_j}{-2 i} \right) \,+\, A^{(4)} \wedge \frac{dw_1 \wedge d\bar{w}_2 + dw_2 \wedge d\bar{w}_1}{-2 i}
\end{equation}
with 
\be
A^{(I)} ~=~{} -\frac{C_{IJK} Z_J Z_K}{2\, Z_3 \cP} \left( dt + \kk \right) + B^{(I)}\,, \qquad~~~ I = 1,\ldots ,4. 
\ee
The $B^{(I)}$ are the magnetic components depending on the coordinates of the four-dimensional base space which can be associated to four magnetic field strengths
\be 
\Theta^{(I)} ~\equiv~{} d_{(4)} B^{(I)} \,,
\ee
where $d_{(4)}$ is the exterior derivative on the four-dimensional base space.
In terms of these quantities, the BPS equations are 
\begin{equation}
\begin{aligned}
\Theta^{(I)} ~&=~{}  \ast_4 \Theta^{(I)}\,, \\
\nabla^2_{(4)} Z_I \:&=\: \frac{1}{2} C_{IJK} \;\! \ast_4 \:\!\! \left(\Theta^{(J)} \wedge \Theta^{(K)} \right), \\
d_{(4)} \kk \:+\: \ast_4 \:\! d_{(4)} \kk \:&=\: Z_I \Theta^{(I)}\,,
\end{aligned}
\end{equation}
where ${\nabla}^2_{(4)} $ is the Laplacian on the four-dimensional base space.

Although supersymmetry significantly simplifies the equations of motion by making them linear on a four-dimensional Euclidean space, they are still complicated to solve. Thus one often makes the further assumption that the hyperk\"{a}hler four-dimensional manifold is a Gibbons-Hawking space. In other words, we assume that the base space has a tri-holomorphic U(1) isometry and an ambipolar hyperk\"{a}hler metric: 
\begin{equation}
ds_4^2 ~=~{}  V^{-1} \! \left( d\psi \,+\,A \right) ^2 \,+\,V ds^2\left(\IR^3 \right), \qquad\nabla^2 V = 0, \qquad  \ast_3 dV\:=\: dA \,, 
\end{equation}
where $ds^2\left(\IR^3 \right)$ is the flat metric on $\IR^3$, and where $d$ is the exterior derivative and $\ast_3$ the Hodge star on $\IR^3$. We denote by $\boldsymbol{\rho}$ the coordinate vector of the three-dimensional base space, so that $ ds^2\left(\IR^3 \right) = d\boldsymbol{\rho}. d\boldsymbol{\rho} $.

The solutions of the BPS equations then take the following form:
\begin{equation}
\begin{aligned}
Z_I ~&=~{} L_I \:+\: \frac{1}{2}C_{IJK} \frac{K^J K^K}{V} \, , \\ 
\kk ~&=~{} \mu \left( d\psi + A \right) \,+\, \varpi   \, , \\
B^{(I)} ~&=~{} \frac{K_I}{V} \left( d\psi + A \right) \,+\, \xi_I,
\end{aligned}
\end{equation}
with 
\begin{equation}
\begin{aligned}
& \nabla^2_{(4)} L_I ~=~{}  0, \qquad \nabla^2 K_I ~=~{}  0, \qquad \ast_3 d\xi_I ~=~ - dK_I \,, \\ 
& \left( \mu \cD V \,-\, V \cD \mu \right) \,+\, \ast_3 \cD \varpi  \,+\, V \partial_\psi \,\varpi ~=~{}  -V \sum_I Z_I \,d \left( V^{-1} K^I\right),
\end{aligned}
\end{equation}
where 
\be
\cD  ~\equiv ~{} d \,-\, A\, \wedge \partial_\psi. 
\ee
By choosing a particular gauge, one can simplify the equation for $\mu$ and $\varpi$,
\begin{equation}
\begin{aligned}
& \mu ~=~{}\frac{1}{6} V^{-2} C_{IJK} K^I K^J K^K \,+\,\frac{1}{2} V^{-1} K^I L_I \,+\, \frac{M}{2} \, , \\ 
&\ast_3 \cD {\varpi}  \,+\, V \partial_\psi \, {\varpi} ~=~{} V \cD M \,-\,  V \cD M \,+\, \sum_I\left(K^I \cD L^I \,-\,  L^I \cD K^I \right), \\
& \nabla^2_{(4)} M ~=~{}  0.
\end{aligned}
\end{equation}
If we further assume that the ansatz quantities are independent of $\psi$, the previous equations simplify considerably and the solutions are uniquely determined by ten harmonic functions on the three-dimensional base space ($V,K^I,L_I,M$). This simplifying assumption is made in Sections \ref{sec:general-limit} and \ref{sec3}.

\subsection{Six-dimensional minimal supergravity coupled to two tensor multiplets}
\label{app:6D}

To write the six-dimensional BPS ansatz in covariant form, we rescale $(Z_4,\Theta_4,G_4) \to (Z_4,\Theta_4,G_4)/\sqrt{2}$, following~\cite{Bena:2017geu}. 
Then we have 
\be
C_{123} ~=~ 1 \,, \qquad C_{344} ~=~ -1 \,. 
\ee
We define the $SO(1,2)$ Minkowski metric via
\bea
\eta_{ab} &=& C_{3ab} \qquad \Rightarrow \qquad \eta_{12} ~=~ \eta_{21} ~=~ 1\,, \quad \eta_{44} = -1 \,,
\eea
which we use to raise and lower $a,b$ indices. After the rescaling of $(Z_4,\Theta_4,G_4)$ we have
\be
\cP ~=~ \coeff{1}{2} \eta^{ab} Z_a Z_b ~=~ Z_1 Z_2 - \coeff12 Z_4^2 \,.
\ee
The metric ansatz \eq{sixmet} is
\begin{equation}
ds_6^2 ~=\, {}   -\frac{2}{\sqrt{\cP}} \, (dv+\beta) \big(dt +  \kk - \tfrac{1}{2}\, Z_3\, (dv+\beta)\big) 
\,+\,  \sqrt{\cP} \, ds_4^2\,. 
\label{sixmet-app}
\end{equation}
The dilaton and axion are given by
\be
e^{2\varphi} \;=\; \frac{Z_1^2}{\cP} \,, \qquad \varsigma \;=\; \frac{Z_4}{Z_1} \,.
\ee
Our ansatz for the tensor fields is
\bea
G^{(a)}  &=&  d_{(4)} \left[ - \frac{1}{2}\,\frac{\eta^{ab} Z_b}{\cal P}\,(du + \kk ) \wedge (dv + \beta)\, \right] ~+~\coeff{1}{2} \;\! \eta^{ab} *_4 \! D Z_b  
~+~ \coeff{1}{2}\,  (dv+ \beta) \wedge \Theta^{(a)} . \qquad~~~
\label{G-ans-cov-app}
\eea
In our conventions, the twisted self-duality condition for the field strengths is 
\be
\ast_6 \:\!\! G^{(a)} ~=~ \T{M}{a}{b} \:\! G^{(b)} \,, \qquad\qquad  M_{ab} ~\equiv~ \frac{Z_a Z_b}{\cP} - \eta_{ab} \,.
\ee

The first layer of the BPS equations then takes the form
\bea
 *_4 D\dot{Z}_a ~=~ & \eta_{ab} D\Theta^{(b)}\,,\qquad D*_4 DZ_a ~=~  - \eta_{ab} \Theta^{(b)} \! \wedge d\beta\,,
\qquad \Theta^{(a)} ~=~ *_4 \Theta^{(a)} .
\eea
The second layer becomes
\begin{equation}
 \begin{aligned}
D \kk + *_4 D\kk -Z_3 \,d\beta 
~=~ & Z_a \Theta^{(a)},  \\ 
 *_4D*_4\!\Bigl(\dot{\kk} +\coeff{1}{2}\,DZ_3 \Bigr) 
~=~& \ddot \cP  -\coeff{1}{2} \eta^{ab} \dot{Z}_a \dot{Z}_b 
-\coeff{1}{4} \eta_{ab} *_4\! \Theta^{(a)}\wedge \Theta^{(b)} .
\end{aligned}
\label{eqFomega-app}
\end{equation} 

\newpage
\section{Analytic solution of the wave equation for large $n$}
\label{app:WEresolution}

In this appendix we describe our method to analytically solve the radial part of the free massless scalar wave equation \eqref{eq:radialeqsegment} in a $(1,0,n)$-superstratum geometry with $n \gg 1$, working to leading order in the $1/n$ expansion. 
From the outset we impose the condition $l \neq -1$ as required by the regularity of the angular wavefunction \eqref{eq:angularconstraint}. For ease of presentation, we consider $p + q_2 \neq 0$. However, one can apply the same method when $p+q_2=0$.

We solve for $K(z)$ subject to
\be
 K(0) \:=\: 0\,, \qquad K(1) \:=\: 0.
\ee
To do so we divide the radial equation in two pieces:
\begin{equation}
\cL\left[\tilde{\omega}\right]  \;\! K(z) \:-\: \frac{z^{n}}{1-z} \;\! \mathcal{E} \left[\tilde{\omega}\right]  \;\! K(z) \:=\: 0 \,, \\
\label{eq:appwaveequ}
\end{equation}
where
\bea
\cL\left[\tilde{\omega}\right]  &\equiv& \partial_z \left(z \,\partial_z\,\right) \:+\: \frac{1}{4(1-z)} \bigg[ \left( p+q_1+\frac{b^2 \tilde{\omega} }{2}\right)^2 - \frac{\left(  p +q_2\right)^2}{z} 
  - \frac{l (l+2)}{ 1-z}  \bigg], \nonumber\\
\mathcal{E} \left[\tilde{\omega}\right]    &\equiv& \frac{b^2 \tilde{\omega} \, \left(4 q_1+b^2\tilde{\omega}\right)}{16} \,,
\eea
where we remind the reader that $q_1$, $q_2$ and $l$ are integers. 

The strategy will be to exploit the fact that $\cL\left[\tilde{\omega}\right] \:\! K = 0$ is analytically solvable, and that the second term in \eq{eq:appwaveequ} can be treated (with some care) as subleading.
We make a series expansion in $1/n^{\nu}$, where $\nu\equiv |l+1|$ was defined in \eq{eq:nu-def},
\begin{equation}
\begin{split}
& K(z) \, =\,  K^{(1)}(z) \, +\, \frac{1}{n^{\nu}} K^{(2)}(z) \,+\,  \frac{1}{n^{2\nu}} K^{(3)}(z) \,+\, \ldots \, , \\
& \tilde{\omega}\, =\,  \omega^{(1)} \, +\, \frac{1}{n^{\nu}} \omega^{(2)} \, +\, \frac{1}{n^{2\nu}} \omega^{(3)} \,+\, \ldots \, .
\label{eq:seriesexpofK}
\end{split}
\end{equation}
The powers of $n$ in this expansion are chosen so that all $\omega^{(j)}$ and $K^{(j)}(z)$ will turn out to be of order one when $n$ is large. 

%
%

We insert this expansion in the wave equation \eqref{eq:appwaveequ}, and we arrange the series expansion in $1/n$ according to our strategy. That is, we put the leading part of the second term in \eq{eq:appwaveequ} on the right-hand side of the second equation below:
\begin{eqnarray}
\cL\,[\omega^{(1)}]  \, K^{(1)}(z) \,  = \, 0\, , \hspace{7.6cm}&&K^{(1)}(0) = K^{(1)}(1) = 0, \nonumber \\
\cL\,[\omega^{(1)} ]   \, K^{(2)}(z) \,= \, \frac{K^{(1)}(z)}{1-z} \left( n^{\nu}\, \mathcal{E}[\omega^{(1)}]   \,z^{n} -b^2 \omega^{(2)}(p+q_1+\frac{b^2}{2} \omega^{(1)})\right), \!\!\!\!\! &&K^{(2)}(0) = K^{(2)}(1) = 0, \nonumber \\
\ldots  \qquad\qquad \qquad \qquad &&
\label{eq:waveeqanalytic}
\end{eqnarray}
and so on at higher order. We will carefully justify this arrangement of terms in what follows. 

If one shows that each $K^{(J)}(z)$ and $\omega^{(J)}$ are of order one when $n$ is large, this guarantees that the series expansion \eqref{eq:seriesexpofK} converges and that the principal features of the solution are captured by $K^{(1)}(z)$. The expansion is similar in spirit to the WKB approximation.

We will need to treat carefully the first term on the second line of~\eq{eq:waveeqanalytic}.
The main subtlety with this term is that even though $z^n \ll 1$ for $z \in [0,1)$, the combination $ \frac{z^{n}}{1-z}$ diverges as $z\rightarrow 1$. However, since $K^{(1)}(z)$ satisfies \eqref{eq:waveeqanalytic} then it behaves as $z\rightarrow 1$ as \eqref{eq:behaviorat1}
\be
(1-z)^{\frac{1+| 1 + l|}{2}} \,.
\ee
Thus for $l\ge1$ and $l\le -3$, the combination $ \frac{z^{n}}{1-z} \,K^{(1)}(z)$ tends to zero as $z \rightarrow 1$.
For $l=0$ and $l=-2$, $ \frac{z^{n}}{1-z} \,K^{(1)}(z)$ tends to a finite value as $z \rightarrow 1$ (although the interval where $ \frac{z^{n}}{1-z} \,K^{(1)}(z)$ is non-negligible is a set of measure zero in the large $n$ limit).
We will carefully analyze the equations and solutions for general $l$ near $z\to 1$ in what follows.

\vspace{-1mm}
\subsubsection*{Derivation of $K^{(1)}$}
\vspace{-2mm}

Let us solve the wave equation \eqref{eq:waveeqanalytic} for $K^{(1)}(z)$ without imposing any boundary condition. There is only one branch of regular solutions for $p + q_2 \in \mathbb{Z}^*$:
\be 
K^{(1)}(z) \: =\: \kappa^{(1)}  \:  z^{\frac{| p +q_2 |}{2}} \:(1-z)^{-\frac{l}{2}}  \: _2 F_1 \left( \frac{\gamma - \delta}{2}\, ,\, \frac{\gamma + \delta}{2} \,  , \, \mu \,,\, z  \right),
\label{eq:generalsol}
\ee
with 
\be
\gamma \: =\: -\, l +| p +q_2 | \,, \qquad
\delta \: =\: p+q_1 +\frac{b^2 \omega}{2} \,, \qquad
\mu \:=\: 1 + | p +q_2 |\,,
\ee
and where $\kappa^{(1)}$ is a constant.
\begin{itemize}
\item Condition $K^{(1)}(1) = 0$.
\end{itemize}
We compute the limit of $K^{(1)}$ \eqref{eq:generalsol} around $z=1$ for the allowed values of $l$:
\begin{align*}
K^{(1)}(z) \underset{z \rightarrow 1}{\sim}  \kappa^{(1)}
\begin{cases} 
\frac{l \,! \:\, \Gamma(\mu)}{\Gamma\left(\mu-\frac{\gamma + \delta}{2}\right)\: \Gamma\left(\mu-\frac{\gamma - \delta}{2}\right)}\:(1-z)^{-\frac{l}{2}} &\:+ \:\:   \underset{z \rightarrow 1}{\cO}\left( (1-z)^{1+\frac{l}{2}}\right) \,, \quad l \geq 0, \vspace{0.5cm} \\ 
%
 \frac{l \,! \:\, \Gamma(\mu)}{\Gamma\left(\frac{\gamma + \delta}{2}\right)\: \Gamma\left(\frac{\gamma - \delta}{2}\right)} \: (1-z)^{1+\frac{l}{2}} &\:+ \:\: \underset{z \rightarrow 1}{\cO}\left( (1-z)^{-\frac{l}{2}} \right) \,, \,\:\quad l \leq -2.
\end{cases}
\end{align*}

We see that the leading-order terms do not tend to zero as $z \to 1$, while the higher-order terms do. Thus, we must set the leading order terms to zero. This can be done by arranging a pole in one of the Gamma functions in the respective denominators:
\begin{alignat}{3}
& l \geq 0 \qquad && \Rightarrow \qquad \Gamma\left(\mu-\frac{\gamma \pm \delta}{2}\right) \,=\, \pm \infty \qquad && \Rightarrow \qquad - \mu +\frac{\gamma \pm \delta}{2} = N \in \mathbb{N}\,,  \nonumber \\
& l \leq -2 \qquad && \Rightarrow \qquad \Gamma\left(\frac{\gamma \pm \delta}{2}\right) \,=\, \pm \infty \qquad && \Rightarrow \qquad -\frac{ \gamma \pm\delta}{2}= N  \in \mathbb{N}\,.
\end{alignat}
Both conditions give the same two towers of permitted values of $\omega^{(1)}$ labelled by $N$,
\begin{equation}
\begin{split}
&\omega^{(1)+}_N \,=\, \frac{2}{b^2} \Big[\,2 N +1+| 1+l | + | p + q_2 | -\left(p+q_1\right) \Big] , \\
&\omega^{(1)-}_N \,=\, -\frac{2}{b^2} \Big[\,2 N +1+| 1+l | + | p + q_2 | +\left(p+q_1\right) \Big] .
\end{split}
\end{equation}
As usual we restrict attention to the positive frequencies, $\omega^{(1)}_N = \omega^{(1)+}_N$. Thus, the radial wave functions are
\begin{equation}
K^{(1)}_N (z) \: = \: \kappa^{(1)}_N \, \left( 1 -z \right)^{\frac{1+| 1+l | }{2}}  \, z^{\frac{| p + q_2 | }{2}} \, \sum_{j=0}^N (-1)^j \, \binom{N}{j} \, \frac{\left(N +1+| 1+l | + | p + q_2 | \right)_j}{\left(1 +| p + q_2 |\right)_j} \, z^j \,.
\label{eq:appwavefunction}
\end{equation}


\begin{itemize}
\item Condition $K^{(1)}(0) =0$.
\end{itemize}
We observe that this condition is automatically satisfied by the radial wave function \eqref{eq:appwavefunction}.

\subsubsection*{Derivation of $K^{(2)}$}
\vspace{-2mm}

From now on, $K^{(1)}$ is fixed to be a polynomial function of the tower labelled by $N$ given in \eqref{eq:appwavefunction}. We must now solve the following differential equation \eqref{eq:waveeqanalytic} to find $ K^{(2)}_N$,
\begin{equation}
\begin{split}
&\cL\,[\omega^{(1)}_N ]   \, K^{(2)}_N(z) \, = \, \frac{K^{(1)}_N(z)}{1-z} \left( n^{\nu}\, \mathcal{E}[\omega^{(1)}_N]   \,z^{n} - \gamma^{(2)}_N\right) , \\
&K_N^{(2)}(0) \;=\; K_N^{(2)}(1) \;=\; 0 \,, 
\label{eq:waveeqanalytic2}
\end{split}
\end{equation}
where for convenience we have defined the constant $\gamma^{(2)}_N$ to be
\begin{equation}
\gamma^{(2)}_N \: \equiv\:b^2 \omega^{(2)}_N \left( p+q_1+\frac{b^2}{2} \omega^{(1)}_N \right) .
\end{equation}
This is the more involved step of the method. It is crucial to show that solutions of \eqref{eq:waveeqanalytic2} do not diverge at the boundaries and are of order one when $n$ is large. If one of these two conditions is not satisfied, the expansion \eqref{eq:seriesexpofK} is ill-defined. We use the standard method of variation of parameters to solve the equation, since we already know that $K^{(1)}_N$ is a solution to the homogeneous equation. We find
\begin{equation}
K_N^{(2)}(z) \:=\: K_N^{(1)}(z) \int_0^z dy \frac{P_N (y) }{y \,  \Big(K_N^{(1)}(y)\Big)^{\! 2}}\,,
\label{eq:KN2firstexpr}
\end{equation}
where $P_N$ is a polynomial defined by
\begin{equation}
P_N (y) \:\equiv\:  \int_0^y dx  \frac{ \left(K_N^{(1)}(x)\right)^2}{1-x}\left( n^{\nu}\, \mathcal{E}[\omega^{(1)}_N]   \,x^{n} - \gamma^{(2)}_N \right).
\label{eq:exprPN}
\end{equation}
At first sight, the integral \eqref{eq:KN2firstexpr} appears likely to be divergent. Indeed, the polynomial $y \left(K^{(1)}_N(y)\right)^{\! 2}$ has $N+2$ distinct roots: a root of multiplicity $| p +q_2|+1 $ at 0,  a root of multiplicity $1+ | 1+l| $ at 1, and $N$ roots of multiplicity 2  between 0 and 1; let us call these intermediate roots $\alpha_j$ for $j=1,\ldots ,N$ (see Eq.\;\eqref{eq:appwavefunction}). We will see that assigning a specific value to $\gamma^{(2)}_N$ will make the function regular and bounded everywhere. Before dealing with the regularity issues at each zero, we first rewrite $\left(K^{(1)}_N(z)\right)^{\! 2}$ in three convenient forms that will be useful in what follows:
\begin{equation}
\begin{split}
\left(K^{(1)}_N(z)\right)^{\! 2} &\:=\:  \kappa^2_N \, \left( 1 -z \right)^{1+| 1+l | }  \, z^{| p + q_2 |} \,  \prod_{j=1}^N (z-\alpha_j)^2 \\
&\:=\:  \kappa^2_N \, z^{| p + q_2 |}\, (1-z) \:\sum_j a_j \,z^j \\
&\:=\:  \kappa^2_N \,\left( 1 -z \right)^{1+| 1+l | } \:\sum_j b_j \,(1-z)^j
\end{split}
\label{eq:formsK1}
\end{equation} 
where $\kappa_N$ is a constant and where the sums on the second and third lines run from 0 to the appropriate maximum values of $j$.
\begin{itemize}
\item \underline{At $z=0$}

The integral \eqref{eq:KN2firstexpr} appears to be ill-defined at $z=0$. However, if we compute $P_N(y)$ with the second formulation of $\left(K^{(1)}_N(z)\right)^2$ in \eqref{eq:formsK1}, the regularity of the integral at 0 is explicit. We integrate \eqref{eq:exprPN}:
\begin{equation}
P_N(y) \:=\: \kappa_N \, y^{| p + q_2 | + 1} \sum_j a_j \left(\frac{n^{\nu}\, \mathcal{E}[\omega^{(1)}_N]}{n + | p + q_2 | + j}   \,y^{n} - \frac{\gamma^{(2)}_N}{| p + q_2 | + j}  \right)\,y^j.
\label{eq:exprofPnat0}
\end{equation}
Moreover, the denominator in \eqref{eq:KN2firstexpr} is
\begin{equation}
y \,  \left(K_N^{(1)}(y)\right)^2 \:=\: \kappa_N \, y^{| p + q_2 |+1} \,  \left( 1 -y \right)^{1+| 1+l | }  \, \prod_{j=0}^N (y-\alpha_j)^2.
\end{equation}
By comparing those two expressions, it is straightforward to see that $\frac{P_N (y) }{y \,  \left(K_N^{(1)}(y)\right)^2}$ takes a finite value at $y=0$ and is integrable at 0. Furthermore, if one takes the limit $z\rightarrow 0$ of the differential equation \eqref{eq:waveeqanalytic2} one can show that $K_N^{(2)}$ has a zero of multiplicity $\frac{|p+q_2|}{2}$ at $z=0$ exactly as $K_N^{(1)}$.
\item \underline{At $z=\alpha_j$}

Obviously, $z=\alpha_j$ is not a zero of $P_N (y)$. So the argument above cannot be used here. However, around $\alpha_j$ we have
\begin{equation}
\begin{split}
&\int_0^z \frac{P_N (y) }{y \,  \left(K_N^{(1)}(y)\right)^2} \:\underset{z\rightarrow \alpha_j}{\sim}\:\int_0^z \frac{dy}{\left(y-\alpha_j\right)^2} \:\underset{z\rightarrow \alpha_j}{\sim}\: \frac{1}{z-\alpha_j},\\
&  K_N^{(1)}(z) \:\underset{z\rightarrow \alpha_j}{\sim}\: \left(z-\alpha_j\right).
\end{split}
\end{equation}
Thus, the product of the two is well-defined and $ K_N^{(2)}$ is well-defined at $z=\alpha_j$.

\item \underline{At $z=1$}

Proving the regularity of $K_N^{(2)}$ around 1 is less straightforward. The most direct argument we found is the following: we compute $P_N(y)$ using the third expression of \eqref{eq:formsK1} and we prove that $P_N(y)$ has a zero of multiplicity $1+\nu$ at 1 for a specific value of $\omega_N^{(2)}$ \eqref{eq:seriesexpofK}. We derive $P_N(y)$ according to \eqref{eq:formsK1}
\begin{equation}
\begin{split}
P_N (y) &\:=\: P_N (1) \,-\, \int_y^1 dx  \frac{ \left(K_N^{(1)}(x)\right)^2}{1-x}\left( n^{\nu}\, \mathcal{E}[\omega^{(1)}_N]   \,x^{n} - \gamma^{(2)}_N \right),\\
& \:=\: P_N (1) \,+\, \left(1-y\right)^{1+\nu}\, \sum_j b_j' \, (1-y)^j.
\end{split}
\end{equation}
where $b_j'$ can be computed from $b_j$ \eqref{eq:formsK1}, $\mathcal{E}[\omega^{(1)}_N]$, $ \gamma^{(2)}_N$ and $n$. Consequently, if we fix $\omega_N^{(2)}$ to satisfy $P_N (1)=0$, $P_N(y)$ has indeed a zero of multiplicity $(1+\nu)$ at 1 which guarantees that $K_N^{(2)}$ takes a finite value. Moreover by expanding \eq{eq:exprofPnat0} in powers of $1/n$ we find 
\begin{equation}
\begin{split}
P_N (y) \:\equiv\:& n^{\nu}\, \mathcal{E}[\omega^{(1)}_N]   \,y^{n+1}   \sum_{\alpha=0} \frac{(-1)^{\alpha}}{n^{\alpha}} (y \partial_y)^\alpha \left( \frac{ \left(K_N^{(1)}(y)\right)^2}{1-y} \right) \\
& - \gamma^{(2)}_N \int_0^y dx  \frac{ \left(K_N^{(1)}(x)\right)^2}{1-x},
\end{split}
\end{equation}
where $(y \partial_y)^\alpha$ means we derive and multiply $\alpha$ times. Evaluating this formula at $y=1$ shows that there always exists a unique solution $\omega_N^{(2)}$ of $P_N (1)=0$ and this value is of order one when $n$ is large. Furthermore, if one takes the limit of the differential equation \eqref{eq:waveeqanalytic2} for this particular value of $\omega_N^{(2)}$ one can show that the finite value $K_N^{(2)}(1)$ must be exactly 0 and the multiplicity of this zero is necessarily $\frac{1+\nu}{2}$ exactly as it is for $K_N^{(1)}$.\\ Finally, $K_N^{(2)}$ does not diverge in $[0,1]$ and it is straightforward from \eqref{eq:exprofPnat0} and \eqref{eq:KN2firstexpr} that $K_N^{(2)}(z)$ is of order one when $n$ is large.

\end{itemize}
In a nutshell, we have shown that the second term of the series expansion \eqref{eq:seriesexpofK} is well-defined: $K_N^{(2)}(z)$ is regular, $K_N^{(2)}(0)=K_N^{(2)}(1)=0$ with the same multiplicity as $K_N^{(1)}$ and $\omega_N^{(2)}$ and $K_N^{(2)}(z)$ are of order one when $n$ is large.

\subsubsection*{Higher-order terms}
\vspace{-2mm}

We finally discuss the higher-order terms of the $\frac{1}{n}$-expansion $\{ K_N^{(3)},\ldots;\omega_N^{(3)},\ldots\}$.
Each term of the expansion satisfies a differential equation of the form
\begin{equation}
\begin{split}
&\cL\,[\omega^{(1)}_N ]   \, K^{(J)}_N(z) \, = \left(\ldots\right) \, K^{(1)}_N(z) + \left(\ldots\right) \, K^{(2)}_N(z) + \ldots + \left(\ldots\right) \, K^{(J-1)}_N(z), \\
&K_N^{(J)}(0) = K_N^{(J)}(1) = 0.
\label{eq:waveeqanalytic3}
\end{split}
\end{equation}
If we know that each function $K^{(K)}_N(z)$ for $K<J$ is well-defined with zeroes at $z=0$ and $z=1$ with multiplicity $\frac{|p+q_2|}{2}$ and $\frac{1+\nu}{2}$ respectively, the same arguments as above can be used to prove that there exists a value for $\omega^{(J)}_N$ where $K^{(J)}_N(z)$ is well-defined with the same kinds of zeroes at $z=0$ and $z=1$.

To conclude, we have solved the wave equation \eqref{eq:WElimitnlarge} in the limit where $n$ is large. We have expanded the solutions for large $n$ and demonstrated the consistency of this expansion. All the features of the tower of solutions are captured by the tower of leading-order terms $K^{(1)}_N$, as discussed in Section \ref{subsec:analyticsolve}.

\end{appendix}

\newpage

\begin{adjustwidth}{-5.5mm}{-5.5mm} 

\bibliographystyle{utphysM}      

\bibliography{microstates}       

\end{adjustwidth}


\end{document}